\title{Translation of Newton's \textit{Principia} into Arabic Under the Aegis of the East India Company: A Rumour Turning into a Myth? }
\author{
        {\em K. Razi Naqvi\/}\\
                Department of Physics,
        Norwegian University of Science and Technology\\
        N-7491 Trondheim, Norway
         }
\date{\today}

\documentclass[11pt,reqno]{article}
\usepackage{latexsym,amssymb,amsmath}
\usepackage[sort&compress,numbers]{natbib}
\usepackage{anyfontsize}
\usepackage{mathtools}
\usepackage{phonetic}
\usepackage{booktabs}
\usepackage{blindtext}
\usepackage{enumitem}
\usepackage{marginnote}
\usepackage{longtable}
\usepackage{soul}
\usepackage{color}
\usepackage{multicol,caption}
\usepackage{multirow}
\usepackage{tipa}%
\usepackage{mathpazo}
\usepackage{arabtex}
\usepackage{utf8}
\usepackage{tikz}
\usepackage{palatino} 
\usepackage{mathpazo}
\usepackage{setspace} 
\usepackage{lettrine}
\usepackage{framed}
\usepackage{fdsymbol}
\usepackage[subnum]{cases}
\usepackage{array}
\usepackage{tabularx}
\makeatletter
\def\hlinewd#1{%
\noalign{\ifnum0=`}\fi\hrule \@height #1 %
\futurelet\reserved@a\@xhline}
\makeatother
\newcommand{\linkcolor}{blue}
\usepackage[colorlinks=true,linkcolor=\linkcolor,urlcolor=\linkcolor]{hyperref} 
\usepackage{url}

\expandafter\def\expandafter\UrlBreaks\expandafter{\UrlBreaks
  \do\a\do\b\do\c\do\d\do\e\do\f\do\g\do\h\do\i\do\j%
  \do\k\do\l\do\m\do\n\do\o\do\p\do\q\do\r\do\s\do\t%
  \do\u\do\v\do\w\do\x\do\y\do\z\do\A\do\B\do\C\do\D%
  \do\E\do\F\do\G\do\H\do\I\do\J\do\K\do\L\do\M\do\N%
  \do\O\do\P\do\Q\do\R\do\S\do\T\do\U\do\V\do\W\do\X%
  \do\Y\do\Z}

\usepackage{titlesec}
\titleformat{\section}
  {\normalfont\fontsize{12}{15}\bfseries}{\thesection}{1em}{}

\titleformat{\appendix}
  {\normalfont\fontsize{12}{15}\bfseries}{\thesection}{1em}{}

\makeatletter
\renewcommand\maketitle
  {\noindent {\setstretch{1.25}
   {\Large\bfseries\@title}\par}%
   \bigskip\par\noindent
   {\large\bfseries\@author}%
   \bigskip\par\noindent
  }
\makeatother
\makeatletter
\DeclareTextCommand{\textprime}{\encodingdefault}{%
\raisebox{-3pt}{\mbox{$\m@th'\kern-\scriptspace$}}%
}
\makeatother

\font\rxii=cmr12

\def\bsq{{\rxii `}}
\def\esq{{\rxii '}}

\usepackage{graphicx}
\newenvironment{Figure}
  {\par\medskip\noindent\minipage{\linewidth}}
  {\endminipage\par\medskip}
\graphicspath{%
    {converted_graphics/}
    {/}
}

\begin{document}

\maketitle

\vspace{2em}
\section*{Abstract}
\label{sec:Abstract}
\vspace{-1ex}
Tafazzul Husain Khan (1727?--1800?), who began his career in the court of Awadh, spent the last two decades of his life as a trusted ally of the East India Company. What set him apart from other court officials was not so much his erudition, political acumen and negotiating prowess, as his favourite pastime: delving into mathematics and astronomy. Contact with the Company personnel, some of whom were conversant with oriental languages and/or contemporary scientific advances, provided him with the opportunity to brush up his mathematical knowledge, and induced him to embark upon---and, according to some, bring to fruition---the task of translating a few important mathematical treatises, among them Newton's \textit{Principia}. According to Campbell, the author of an obituary notice (published in 1804), ``he translated the Principia from the original Latin, into Arabic''. The evidence gathered by Campbell is examined, and found insufficient to warrant this astounding and oft-repeated claim. Of the three tracts authored by Tafazzul (all published posthumously in abridged versions), none can be described as a translation of Newton's \textit{Principia}. Until the emergence of some tangible evidence, any talk of his translations of the \textit{Principia} and other western treatises can only be characterised as rumour, a process in which recall is often accompanied by distortion.\\[1em]
\textit{Keywords}: Indian Mathematics (1780--1830), Reuben Burrow, Tafazzul Husain Khan,  John Tytler, Diwan Kanh Ji, Maulavi Ghulam Husain\\[2em]

\begin{multicols}{2}
\section{Introduction}
\label{sec:intro}
``Don't cite a publication unless you have read it yourself'' is one of the golden rules of academic authorship \cite{Blanchard1974RefUnref}, and perhaps one that is flouted most frequently. But even those who abide by the rule merely shift the burden of veracity from \textit{their} shoulders to those of the author(s) they cite. If this primary source happens to be reliable, and is cited by a multitude of subsequent authors, the process serves to advance learning. However, if the primary source is of doubtful authenticity, or  contains factual errors---even small ones (for example, incorrect dates, erroneous spellings of proper names)---multiple citations of this work will generate what may be called the academic equivalent of rumour.

To offer a justification for invoking the concept of \textit{rumour}, I begin with the words of Bernard Hart, a British psychiatrist \cite{Hart1916Rumour}:

\medskip
\leftskip=2em
{\small
Rumour is a complex phenomenon consisting essentially in the transmission of a report through a succession of individuals. It may be provisionally regarded as the product of a series of witnesses, each of whom bears testimony to a statement imparted to him by his predecessor in the series. The reliability of a rumour depends, therefore, upon the accuracy with which each such statement is transmitted, and ultimately upon the accuracy of the report furnished by the first member of the series, who is assumed actually to have seen or heard the event in question. 
\par }
\medskip
\leftskip = 0em

To be sure, the above passage refers specifically to information communicated by word of mouth, but some trivial adjustments will make it applicable to formal citations or written statements about the authorship of manuscripts.

Three decades later two American psychologists, Allport and Postman (A\&P), wrote an entire book on the psychology of rumour \cite{Allport1948Psychology}.
They defined a rumour as a specific (or topical) proposition for belief, passed along from person to person, usually by word of mouth, without secure standards of evidence being present. The central feature of their definition is its insistence that rumour thrives only in the \textit{absence of secure standards of evidence}. A\&P identified two prerequisites for rumour:  the theme of the story must be of some import to both the speaker and listener, and the true facts must be shrouded in some kind of ambiguity. Among the multiple causes for this ambiguity, and the resulting (often involuntary) corruption of the original report, only one---the incapacity of the rumour-receiver to grasp the vital detail(s)---is germane to the issue discussed below.

\section{A note on transliteration}
\label{sec:transliter}

In the main body of the text, romanization of Persian and Urdu words will be carried out by supplementing the familiar substitutions with three diacritical signs: a macron (a bar) placed over a short vowel sign will indicate the lengthening of the vowel, and the following substitutions will be used: {\seturdu\<`>} = \bsq\ and {\seturdu\novocalize{\<"'>}} (\textit{hamza}) = \esq. A more elaborate notation, explained in Appendix~\ref{appendix:translit}, will be used in the bibliography.

\section{Who was Tafazzul Husain Kh\={a}n?}
\label{sec:whowas}

Tafazzul Husain Kh\={a}n (1727?--1801?), who will henceforth be called Tafazzul, was known in his days as a litt\'{e}rateur, mathematician, diplomat and much else. He lived an eventful life, earning both praise and opprobrium, but \textit{here} only those events will be highlighted  which are relevant to his academic activities. The qualifier \textit{Kashm\={\i}r\={\i}} was often appended to the above three names, but he was born in Sialkot, Kashmir being the region where his ancestors had once lived \cite{GhulamAli1864Imad}. The honorific, \textit{Kh\={a}n-i \bsq All\={a}m\={a}}, meaning ``a scholar par excellence'' became a part of his name after he achieved eminence for his multisided erudition. 

\section{Principal sources for Tafazzul's life and works}
\label{sec:sources}
The standard reference for the personal and political aspects of Tafazzul's life is \textit{\bsq Im\={a}d al-Sa\bsq\={a}dat} \cite{GhulamAli1864Imad}, written by Ghul\={a}m \bsq Al\={\i} Kh\={a}n, an employee of the East India Company.  Among other frequently cited authors in this context are Mirz\={a} Ab\={u} T\={a}leb\cite{Hoey1885AsafuD}, Basu \cite{Basu1943Oudh} and Cole \cite{Cole1988Roots}. A short but informative biographical sketch is available in Guenther's article \cite{Guenther2010Fontana}.

The major sources dealing with various aspects of Tafazzul's academic interests and activities are listed below. Each source will be assigned a label (written in bold italics) which will be used for further reference to it in the rest of this article. \\

{\small
\noindent \textbf{\textit{Tuhfa}}: A Persian book \cite{Shushtari1847Tuhfa} with a title usually abbreviated as \textit{Tuhfat al-\bsq{\=A}lam}. Authored by \bsq Abd al-Lat\={\i}f Kh\={a}n Sh\={u}shtar\={\i}, who became a personal friend of Tafazzul, this book provides glimpses into the scholarly schedule followed by Tafazzul during the last years of his life.\\[1em]
\noindent \textbf{\textit{Obituary}}: An obituary notice  written by Lawrence Dundas Campbell \cite{Campbell1804AARegister}. More details about this source are given later.\\[1em]
\noindent \textbf{\textit{Leaflet}}: Syed Mahomed Ali (hereafter SMA), a descendant of Tafazzul, published a leaflet titled \textit{Life of Tuffuzzool Hussain Khan}, a choice that might lead a reader into expecting more than is delivered \cite{KhanSMA1908TafazLife}; it consists of five \textit{Extracts} (1--5, listed on pp.~i--ii), none of which is from the pen of the compiler. Part I consists of \textit{Extracts} 1 and 2, the first of which is taken from Lord Teignmouth's \textit{Memoir} published in 1843 \cite{Teignmouth1843Memoir}, and the second  from a review of the \textit{Memoir}, published a year later \cite{CalRev1844Teign}. Part III, which is of no interest to us, reproduces \textit{Extracts} 3 and 5, two official letters concerning pensions granted by the Company to Tafazzul's cousin and son. The text of Part II was meant to be identical with that of \textbf{\textit{Obituary}}, but a great many clerical discrepancies and errors crept in when the text was typeset for inclusion in \textbf{\textit{Leaflet}} (see Appendix~\ref{appendix:ScrutPamp}).  The original sources for the texts of \textit{Extracts} 1, 2 and 4 have existed in the public domain for quite some time, which means that citing \textbf{\textit{Leaflet}} now amounts to drinking polluted water when clean is available.\\[1em]
\noindent \textbf{\textit{UrduBio}}:   Some seven years after the publication of \textbf{\textit{Leaflet}}, SMA wrote  an Urdu biography \cite{KhanSMA1915TafazBio}, which draws from Ref.~\cite{GhulamAli1864Imad}, \textbf{\textit{Tuhfa}} and \textbf{\textit{Obituary}}, and provides very little additional information of value. \\[1em]
\noindent \textbf{\textit{Chronograms}}:  Thomas William Beale, whose main literary interest was collection and composition of chronograms (in Persian and Urdu), published a massive collection of chronograms (along with some biographical information in prose) under the title \textit{Mift\={a}h al-Taw\={a}r\={\i}kh} \cite{Beale1867Miftah}. This rather unusual compilation  (in Persian) was described in some detail by Elliot in \textit{The History of India as told by Its Own Historians, The Muhammadan Period}  \cite{Elliot1877History}. The literal meaning of the title is \textit{Key to Histories}, but a glance at the contents makes it clear that Beale is using the noun \textit{t\={a}r\={\i}kh}, which could mean \textit{history} or  \textit{chronogram}, in both senses. A concise biographical note, not a single word of which is superfluous, may be found in this book. A translation of the note will be presented later. 
\par }

\smallskip
So far as Tafazzul's mathematical studies and writings are concerned, there are only two cardinal references, namely \textbf{\textit{Tuhfa}} and \textbf{\textit{Obituary}}. When allowance is made for Sh\={u}shtar\={\i}'s numerous mistransliterations of European names (see below), \textbf{\textit{Tuhfa}}  and \textbf{\textit{Obituary}} agree except on one crucial point: \textbf{\textit{Tuhfa}} does not include the \textit{Principia} among the books translated by Tafazzul.

A minute examination of these sources will not be carried out in this section, but some general observations appear to be necessary for preparing the ground. 

It is difficult enough to transcribe European names in Urdu, but the task becomes much harder for an author writing in Persian, which has a significantly smaller inventory of consonants and vowels than Urdu; furthermore, Sh\={u}shtar\={\i} follows (with very few exceptions) the customary omission of short vowels even when he transcribes European names, and seems to rely on memory rather than meticulous note-taking. To take just one example now, he mentions a certain {\seturdu\novocalize\<mistar bArlw>}, calls him ``a sage the likes of whom are few even in England'' \cite[p.~454]{Shushtari1847Tuhfa}, and states that it was this scholar who imparted Western learning to Tafazzul.  When Edward Rehatsek (1819--91) reviewed \textbf{\textit{Tuhfa}} for a library catalogue edited by him \cite{Rehatsek1873Firuz}, he came to the eminently reasonable conclusion that Sh\={u}shtar\={\i} must have meant ``Mr. Barlow'', and there indeed was a Sir George Hilaro Barlow in the neighbourhood \cite{Buckland1906IndBio}, but hardly likely to have been a good enough mathematician to be called a sage with few peers. The name of the real sage was ``Reuben Burrow'' \cite{Buckland1906IndBio}. Other spelling aberrations will be mentioned later. 

Far more alarming than transcription idiosyncrasies and scribal transgressions is that Sh\={u}shtar\={\i} has earned the reputation of having been well informed about post-Newton astronomy. For example, Schaffer writes \cite{Schaffer2009Astro}: ``\textrevglotstop Abdu\textglotstop l-La\d{t}\={\i}f Shushtar\={\i}, Tafazzul's friend and biographer, \ldots\ learnt the orthodox Newtonian views that comets were planets moving in ellipses round the Sun in one focus''. This is what \textit{we} have learnt about planets and periodic comets from Kepler and Newton. As to what Sh\={u}shtar\={\i} was told by his astronomically educated friends we can only speculate, but anyone who reads \textbf{\textit{Tuhfa}} will find out that, according to its author, the sun is located at the centre of the ellipse! Even those who cannot read Persian will be able to see his ``heliocentric'' illustration of the solar system on p.~360. Since the lithographic edition was published long after the author's death, it is important to rule out the possibility that the illustration (in which Mars and Earth are shown, probably as a result of an oversight, orbiting along a common path!) was prepared by an inattentive person who atrociously misrepresented the trajectory of a comet drawn by the author, one should go to p.~352 (line 2) and note the word {\seturdu\novocalize\<was.t>}, which means ``in middle or centre of''.
\begin{Figure}
 \centering
 \includegraphics[width=\linewidth]{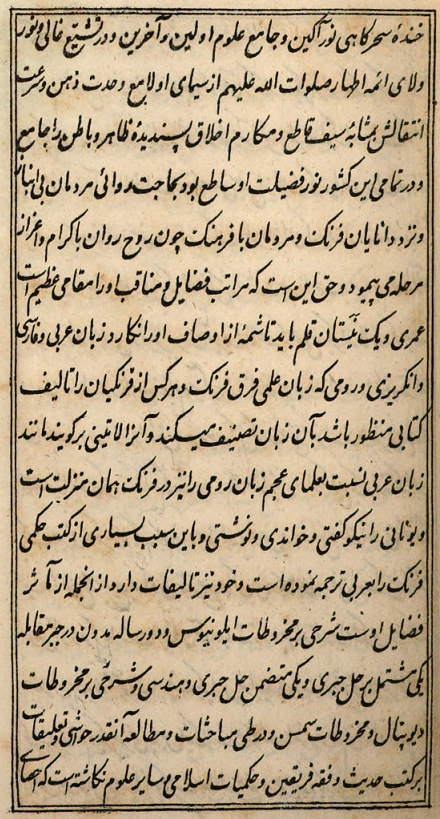}
 \captionof{figure}{Page 443 of \textbf{\textit{Tuhfa}} \cite[p.~443]{Shushtari1847Tuhfa}; an English translation of the text is presented in Appendix~\ref{appendix:TafBio}.}
\label{fig:1}
\end{Figure}
A reader of \textbf{\textit{Tuhfa}} with a sound knowledge of elementary astronomy cannot fail to notice that its author had misconstrued the teachings of Newton. Also, our author appears to be out of touch with the history of astronomy, for he calls (p.~351) {\seturdu\novocalize\<kwparnikws>} (\textit{k\={u}parnik\={u}s}$=$Copernicus), the inventor of the telescope!  One is driven to the uncharitable conclusion that, despite claims to the contrary, Sh\={u}shtar\={\i}'s knowledge of European astronomy and mathematics was much too paltry and muddled to earn him a place in this discussion.

\section{\mbox{The received wisdom about Tafazzul.}  Part 1}
\label{sec:recdwisd1}

For describing Tafazzul's mathematical exertions and their concrete manifestations, Rizvi, the author of a two-volume work on the socio-intellectual history of Twelver Sh\={\i}\bsq ism in India \cite{Rizvi1986Socio}, has essentially paraphrased the account given on p.~443 of \textbf{\textit{Tuhfa}} (see Figure~\ref{fig:1}). Most other authors have relied heavily, and a few exclusively, on the information presented in \textbf{\textit{Obituary}}. \par

In the rest of this section, braces are used to separate my passing comments on Rizvi's account \cite[p.~228]{Rizvi1986Socio}, which is reproduced below (with all the typographic errors left intact): \par

\smallskip
\leftskip=2em
{\small\noindent
\ldots Tafazzul Husayn learnt Greek, Latin and English and obtained considerable proficiency in these languages. He translated many philosophical works from Western languages into Arabic and wrote some original ones on philosophy, \textit{hikma} and mathematics. He was the author of the following works: 
\begin{enumerate}[font=\itshape]
\item Commentary on the \textit{makhr\={u}tat} (Conica) of Abull\={u}niy\={u}s (Appollonus) of Tyana (ca 81--96).\\{\footnotesize\{It will become clear, after reading \S~\ref{sec:Hastings}, that this notion is probably based on a misunderstanding of the contents of the manuscripts Tafazzul \textit{copied} for Warren Hastings. As for the phrase ``of Tyana'' (inserted by Rizvi), let us note that \textit{our} Apollonius, the famous geometer, was from Perga, and was born some 250 years before Jesus; his namesake from Tyana will be recalled later for a different purpose.\}}
\item Two treatises on Algebra. \\{\footnotesize\{Sh\={u}shtar\={\i} stated: ``two treatises on algebra, one containing algebraic solutions, the other algebro-geometric solutions''. We will return to this item in \S~\ref{sec:3MSS}.\} }
\item Commentary on the \textit{makhr\={u}tat} by Devanpal [Diophant and Simson/Robert Simson].\\{\footnotesize\{The text within the square brackets, added by Rizvi, is his decipherment of the two names mentioned in Sh\={u}shtar\={\i}'s text. Recall that short vowels are usually omitted in Persian writing. The second name has no long vowels and its consonants are \textit{smsn}. Rizvi inferred that the cluster stood for \textit{simsan}, written in Persian characters as {\seturdu\novocalize\<simsan>}; Robert Simson, a recognised mathematician, did write a text on conics, but Tafazzul made no mention of Simson, not for this book nor for any other by him. As for the first author, readers familiar with the names of European mathematicians would be able to guess that they are looking at a mauled form of \textit{de l'Hospital}, the French mathematician whose name is familiar to every student of calculus.  With this background, Sh\={u}shtar\={\i}'s text can be given the following truly literal rendering: ``commentary on conics of \textit{delopit\={a}l} and on conics of \textit{simsan}''.\}}
\item Persian translation of Newton's (d. 1827) \textit{Philosophiac naturalis principiamathematica.}\\{\footnotesize\{\textit{Persian} translation?  In lines 8--10 of the text in Figure~\ref{fig:1}, Sh\={u}shtar\={\i} likens the status of Latin in Europe to that of Arabic in the non-Arab Muslim world, and speaks specifically of translations (by Tafazzul) of several European philosophical works into \textit{Arabic}. In fact, Rizvi himself states, in the second sentence of the passage quoted above ``translated \ldots \ldots from Western languages into Arabic''. The ``Arabic or Persian?'' question will continue to vex us. (Newton died in 1727.)\}}

\item A book on Physics. \\{\footnotesize\{This item and the next have no counterparts in  Sh\={u}shtar\={\i}'s text.\}}
\item A book on Western astronomy. \{See \S~\ref{sec:3MSS}.\}
\end{enumerate}
\par }
\medskip
\leftskip = 0em

It is natural to enquire into the fate of Tafazzul's putative writings (original works as well as translations). Rizvi, having anticipated the question, does not keep his readers waiting, and answers it immediately after his list of Tafazzul's written contributions:

\medskip
\leftskip=2em
{\small
Some of these books [which ones?] were taught in Sh\={\i}\bsq\={\i} seminaries in the nineteenth century but are now scarce. He also wrote commentaries and glosses on the works of \textit{fiqh}. His devotion to teaching and studies knew no bounds. Early in the morning he taught mathematics to scholars. He then performed his official duties. In the afternoon he lectured on Im\={a}miyya (Isn\={a} \textrevglotstop Ashariyya) \textit{fiqh}. Before sunset he taught Hanafi \textit{fiqh}. After night prayers immersed himself in study and research. After his morning prayers he slept for a very short time. Before he went to bed his musicians played for him. No physician could persuade him to take more rest. He was enamoured of the company of scholars. Shustar\={\i} [Rizvi's spelling] frequently called on Tafazzul Husayn. The latter also paid return visits and both discussed problems of rational and traditional sciences. Shustar\={\i} was proud of considering himself as one of Tafazzul Husayn's disciple[s], although he had not studied regularly under him. 
\par }
\medskip
\leftskip = 0em

Apart from the first sentence (which looks, in the absence of concrete evidence, a face-saving gesture), the above passage is essentially a translation of what Sh\={u}shtar\={\i} wrote in \textbf{\textit{Tuhfa}}  \cite[p.~444]{Shushtari1847Tuhfa}. I cannot help recall what Gibbon wrote about Apollonius of Tyana \cite{Gibbon1887Rome}: ``His life is related in so fabulous a manner by his disciples, that we are at a loss to discover whether he was a sage, an imposter, or a fanatic.'' The last epithet will have to be replaced, in Tafazzul's case, by one of the many disparaging alternatives used by his detractors \cite{Beale1867Miftah,Aziz1896Malfuzat,Bhanu1979Agra}: utter unbeliever, perfidious and treacherous, spy.

\begin{Figure}
 \centering
 \includegraphics[width=\linewidth]{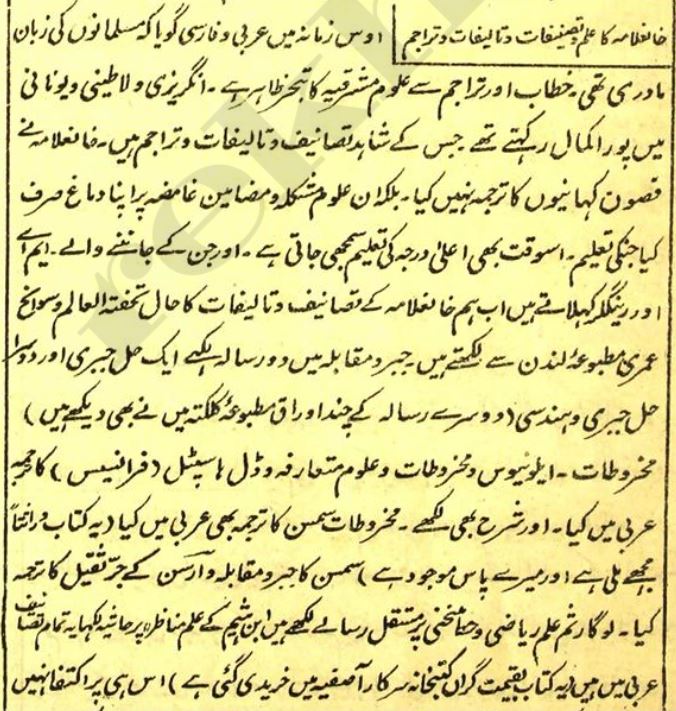}
 \captionof{figure}{An excerpt from \textbf{\textit{UrduBio}} \cite[p.~37]{KhanSMA1915TafazBio}.}
\label{fig:2}
\end{Figure}  

Let us turn now to \textbf{\textit{UrduBio}} \cite{KhanSMA1915TafazBio}. The account of Tafazzul's scholarly output occupies about a page and a half in this booklet with fifty pages of text; an excerpt of the relevant section is displayed in Figure~\ref{fig:2}. Urdu authors of that era (early twentieth century) used few punctuation marks, and made excessive use of the letter ``{\seturdu\novocalize\<w>}'' for a purpose similar to that served by the ampersand sign ``\&''; in Figure~\ref{fig:2}, one sees only dashes, which are Urdu equivalents for full stops, and parentheses. SMA uses  Sh\={u}shtar\={\i}'s incorrect, two-dotted spelling {\seturdu\novocalize\<aylwniyws>} for Apollonius, writes Emerson as  {\seturdu\novocalize\<irsan>} (\textit{Irson}), and  de l'Hospital as  {\seturdu\novocalize\<,dil ,hAspi,tal>}, but in the literal translation presented below, such slips will be ignored; a few additional pauses will also be inserted for the sake of making the passage readable.
 
After extolling Tafazzul's mastery of Arabic and Persian, SMA tells us that he enjoyed 

\medskip
\leftskip=2em
{\small\noindent 
complete command over English \& Latin \& Greek, the proof of which are his writings \& compilations \& translations. Kh\={a}n-i \bsq All\={a}ma did not translate romances \& fables; rather, he taxed his mind with hard sciences \& abstruse topics of the kind whose study is considered even today as advanced education, and those who acquire knowledge of this kind are called M.A.'s and wranglers. We will now refer to \textit{Tuhfat al-\bsq{\=A}lam} [\textbf{\textit{Tuhfa}}] and the biography published in London [\textbf{\textit{Obituary}}] for describing Kh\={a}n-i \bsq All\={a}ma's (original) compositions \& compilations. [He] wrote two algebraic treatises, one on algebraic solutions and another on algebraic \& geometric solutions. (Indeed, I have seen some pages of the second treatise, published in Calcutta). [He] translated into Arabic Apollonius's conics \& conics \& common notions [alternatively, axioms] \& de l'Hospital (Frenchman) and also wrote commentaries. [He] also translated Simson's conics into Arabic (I inherited this book and it is still in my possession).  [He] translated  Simson's algebra and Emerson's mechanics. [He] wrote \textit{mustaqil} tracts [\textit{tracts of lasting value}?] on logarithms, mathematical science \& curves, marginal comments on Ibn-i Haytham's knowledge (or science) of disputations [see below]. All these works are  in Arabic. (This book was acquired at great expense by the Government \={A}sifya Library. This is not all. [Figure~\ref{fig:2} ends here; Tafazzul's theological writings are mentioned in the last two lines of the page, and excerpts from \textbf{\textit{Obituary}} are quoted on the next page.]
\par }
\medskip
\leftskip = 0em

Unfortunately, one cannot tell which particular book was bought by the library of the Hyderabad State. It is also regrettable that SMA could not arrange for a publication of the translation of Simson's book on conics, making it available for examination by those who believe that seeing (but not only seeing, also reading carefully) is believing. What engulfs SMA's claim (about inheriting the translation of Simson's \textit{Conics}) in a dense cloud of doubt, calling into question his competence, is the fact that Tafazzul (who never mentioned Simson) claimed to have translated, among other books, Thomas \textit{Simpson}'s book on \textit{algebra} and a tract of Apollonius (but \textit{not} that on conic sections).

The phrase ``knowledge (or science) of disputations'' is my translation of the title {\seturdu\novocalize\<`ilm munA.ziraH>} mentioned by SMA; if ibn al-Haytham (Latinized as Alhazen) did write such a book, I have yet to come across it. I will assume that SMA meant  {\seturdu\novocalize\<kitAb almunA.zir>} (\textit{Kit\={a}b al-man\={a}\.{z}ir}), usually translated as \textit{Book of Optics}. 

A concise summary of what we have just read, unenucmbered by cavillings on SMA's faulty text, would be helpful before we move on. SMA averred that he would base his account on \textbf{\textit{Tuhfa}} and \textbf{\textit{Obituary}}, but the text itself does not bear this out. Let us list some of the discrepancies.
\begin{enumerate}
\item
Whereas \textbf{\textit{Tuhfa}} states that Tafazzul wrote a \textit{commentary} on Apollonius's \textit{Conics}, SMA claims that Tafazzul \textit{translated} this book into Arabic. \textbf{\textit{Obituary}} speaks neither of a commentary nor of a translation. 
\item
\textbf   SMA chooses to include the two books on algebra mentioned in \textbf{\textit{Tuhfa}}, but not in \textbf{\textit{Obituary}}; he also claims to have seen some pages of one of these books.
\item
Neither \textbf{\textit{Obituary}} nor \textbf{\textit{Tuhfa}} credits Tafazzul with marginal comments (glosses) on Alhazen's \textit{Book of Optics}.
\end{enumerate}
Let us conclude this section by looking at what Beale wrote in \textbf{\textit{Chronograms}} \cite{Beale1867Miftah}. The prose part of his note is translated below:

\medskip
\leftskip=2em
{\small 
\begin{center}
\textbf{Tafazzul Husain Kh\={a}n Kashm\={\i}r\={\i}}
\end{center}

He is also known as Kh\={a}n-i \bsq All\={a}ma. Among all the writings of this peerless individual are one on the astronomy of the \textit{\d{h}ukam\={a}} (philosophers) of Europe and two more manuscripts on \textit{\d{s}an\={a}\bsq at-i jabr wa muq\={a}bala} (the art of algebra). Shortly before the demise of Naw\={a}b \={A}\d{s}af-ud Daula he [Tafazzul] achieved eminence as the Naw\={a}b's representative; afterwards, during the reign of Naw\={a}b Sa\bsq\={a}dat Al\={\i} Kh\={a}n, he went for a sojourn in Calcutta; while returning home, he passed away in Murshid-\={A}b\={a}d on the fifteenth of Shaww\={a}l 1215 Hijra [which corresponds, according to my reckoning, to 1st March 1801]. The text of a panegyrical chronogram penned by Sh\={a}h Mu\d{h}ammad Ajmal Il\={a}h-\={A}b\={a}d\={\i} is given below [skipped here]. \par }
\medskip
\leftskip = 0em

The reader will have noticed that Beale does not mention \textit{any} works of translation. He refers to only three books, which correspond to items {\it 6} and {\it 2} in Rizvi's list (at the beginning of \S~\ref{sec:recdwisd1}). It will be convenient to introduce descriptive names for these books; accordingly I will name item {\it 6} as \textbf{\textit{Copernican Astronomy}}, and the other two tracts as \textbf{\textit{Algebra}}  and \textbf{\textit{Algebra-in-Geometry}}. 

\medskip
\section{\mbox{The received wisdom about Tafazzul.} Part 2}
\label{sec:recdwisd2}

The news that Tafazzul had accomplished an Arabic translation of Newton's \textit{Principia} was broken to the English-reading public by the publication of \textbf{\textit{Obituary}}, an essay written by Lawrence Dundas Campbell  \cite{Campbell1804AARegister}, the then editor of a book series with a title that will be abbreviated in the main text (but not in the bibliography) as \textit{Asiatic Annual Register}. Successive volumes in the series are not numbered, and the editor's name appears in the front matter of only some volumes; however, the missing information may be found in a catalogue that is available in the public domain \cite{AAReg-HathiTDL}. For our purpose it will be sufficient to note that every volume is divided into many independent sections, each of which is paginated separately, and that biographical accounts of various notables, including Tafazzul's  obituary, are presented in a section designated as ``Characters''.

The subscribers to the  \textit{Asiatic Annual Register} (who are listed at the beginning of each volume) were deeply interested in the affairs of the East India Company, and many, perhaps most, of them must have been familiar with the names of the top ranking servants of the Company (and their immediate subordinates). The likes of Warren Hastings, Lord Teignmouth, Marquis Cornwallis and Sir William Jones still need no introduction, but the reader of this article would find it helpful to acquire a nodding acquaintance with some other characters, including two brothers,  David Anderson (1750--1828) and James Anderson (1758--1833), who are commemorated in the gallery of benefactors of Edinburgh University Library. We are told \cite{AndersonBrosEdinUniv} that they may have both studied 

\medskip
\leftskip=2em
{\small\noindent
at the University of Edinburgh like their elder brother Francis, but only James appears to have graduated. They entered the service of the East India Company, David as a writer or clerk, and James as a cadet in the HEIC army. They became assistants to and close friends of Warren Hastings, Governor-General of Bengal, for whom David was a major political diplomat, and James a Persian interpreter. David returned to England with Hastings in 1785, and gave evidence for the defence at Hastings' impeachment; James returned to England the following year. David helped Hastings prepare his defence for his impeachment, and was one of the few witnesses who refused to be browbeaten by the managers of the prosecution, Edmund Burke, Charles James Fox and Richard Brinsley Sheridan.

Like Hastings they assembled their own collections of Oriental books and manuscripts. David gifted 113 volumes from his extensive collections of Oriental manuscripts to the University, and James' nephew Adam Anderson gifted his uncle's 54 Persian manuscripts after the latter's death. \par }
\medskip
\leftskip = 0em

\section{Synopsis of Tafazzul's English obituary}
\label{sec:synopobit}

The text of \textbf{\textit{Obituary}} occupies eight double-column pages. Campbell informs us that he solicited information about Tafazzul's life and works from David Anderson and Lord Teignmouth, and both replied. The former also enclosed two letters he had received, after leaving India, from Tafazzul. When a passage is quoted from \textbf{\textit{Obituary}}, the spellings for proper names written in Latin characters will be retained; three of the excerpts reproduced here contain a footnote ({\bf FN}) each, which are placed at the end of the quoted passage. 

The first three columns of \textbf{\textit{Obituary}} contain introductory comments and a brief account of Tafazzul's life from his birth ``in the celebrated valley of Cashmir'' [!]  up to the time when he decided to leave the court of Lucknow and accept an offer from Warren Hastings to become ``assistant to Major Palmer in conducting some political negotiations with the Rana of Gohud''. For subsequent events in Tafazzul's life, we turn to David Anderson's letter, which was reproduced in toto by Campbell, but here 
we will be content with a single paragraph:

\medskip
\leftskip=2em
{\small\noindent

During the intervals of these tedious and vexatious negotiations [in November 1781], Tofuzzel Hussein delighted to pass his time with my brother, Mr. Blaine, and myself, in conversing on the different laws, customs, and manners of Europe and of Asia; on Persic, Arabic, and Hindu literature; and above all, on the sciences of mathematics and astronomy, in which he had made a considerable proficiency, derived partly from his study of Arabian authors, and partly from his communications with \textit{the learned Mr. Broome} [my italics]. These conversations he always enlivened, by occasionally intermixing sallies of wit and pleasantry. He became, at this time, anxious to learn the English language, and my brother took great pains to teach it to him. He did not then make much progress, but he continued to pursue this study with such ardour and application, that he was, some years afterwards, able, not only to read, but to write English with accuracy. \par }

\medskip
\leftskip = 0em

Some credible evidence, to be presented in  Section~\ref{sec:Hastings}, suggests that in all likelihood ``the learned Mr. Broome'' was Captain Ralph Broome. Reuben Burrow, the man who had the capacity to widen Tafazzul's mathematical horizons, was still in England in 1781. In \textbf{\textit{Obituary}} his name is consistently misspelt as ``Ruben Burrows'', and the error has percolated to the works of those who have cited this source without checking whether a mathematician with such a name went to work in India in late 18th century.

After Anderson's letter (which occupies just over four columns) comes to its end, Campbell continues the narrative in his own words:

\medskip
\leftskip=2em
{\small\noindent
In 1788, a reconciliation took place between the vizier Assof-ud-Dowlah and Tofuzzel Hussein, and the latter was soon after appointed vakeel from the court of Lucknow to the British government. In this capacity he resided some years at Calcutta, where he cultivated the society of Sir William Jones and Lord Teignmouth (then Mr. Shore), and where, at the hospitable mansion of his friend Mr. Richard Johnson, at Russipughilee, he had every facility afforded him of pursuing his favourite studies of mathematics and astronomy; and had also an opportunity of availing himself of the instruction of Mr. Ruben Burrows, the celebrated mathematician; by which means he acquired a knowledge of the philosophy of Newton. And with a view of combining his study of the languages with that of the sciences, he translated the Principia from the original Latin, into Arabic. 
 \par }

\medskip
\leftskip = 0em

The material presented immediately afterwards (four columns, amounting to almost a quarter of the entire text) quotes at length from the two letters sent by Tafazzul to David Anderson, in both of which he speaks mainly of his services in the interests of the Company, but one paragraph in the second letter (written in Persian) does provide a glimpse of his scholarly activities:

\medskip
\leftskip=2em
{\small\noindent
You ask me if I continue my studies as usual, or if my employment in public business has diverted my thoughts from literary pursuits?---Some time ago, I employed myself, for a few months, in reading the history of England, chiefly with a view of acquiring competent knowledge of the language. I have since given it up, and have been engaged in translating the Principia of Sir Isaac Newton, Thomas Simpson's book on Algebra, Emerson on Mechanics, Appolonius de \textit{Sectione Rationis}, translated into latin by doctor Halley, and a work on Conic Sections by ({\seturdu\novocalize\<dAlwptAl>})${}^\ast$ \textit{Del-hopital} a Frenchman. All these books I am translating into Arabic, besides several short treaties on Logarithems, curve lines, \&c. \&c. Some of them I have already finished, and some more of them will soon be brought to a conclusion.---In short, I continue to devote my leisure hours to these pursuits.\\ \par
\vspace{-1em}
\noindent------------\\ 
\noindent \noindent {\bf FN}: ${}^\ast$ \textit{Del-hospital}. William Francis, Marquis de 'l Hospital, the celebrated author of the L'Analyse des infinimens Petits, and the friend of Malbranche. 
\par }

\medskip
\leftskip = 0em

The text of this footnote has so many blemishes that a correction is warranted: ``Guillaume Fran\c{c}ois Antoine de l'Hospital, the celebrated author of the \textit{Analyse des infiniment petits}, and the friend of Malebranche.'' 

In \textbf{\textit{Obituary}} Campbell corroborates Tafazzul's account of his ``literary pursuits'' by adducing an ``extract of a letter from his friend and associate in these labours, Mr. Ruben Burrows, to Lord Teignmouth''. The excerpt reads:

\medskip
\leftskip=2em
{\small\noindent
Tofuzzel Hussein continues translating the Principia of Newton, and I think we shall soon begin to print it here in Arabic:---my notes and explanations are to accompany the translation$\dagger$.---He has likewise translated Emerson's Mechanics, and a Treatise on Algebra, (that I wrote for him) into Arabic. He is now employed in translating Apollonius de Sectione Rationis. The fate of this work is singular; it was translated from Greek into Arabic, and the Greek original was lost; it was afterwards translated from Arabic into Latin, from an old manuscript in the Bodleian library; the Arabic of it is now totally lost in Asia.---\textit{I translated the Latin version into English, and from the English Tofuzzel Hussein is now rendering it into Arabic again}. [My italics]\\ \par
\vspace{-1em}
\noindent------------\\ 
\noindent \noindent {\bf FN}: $\dagger$ The translation was finished, but it has not been printed; and we believe Mr. Burrows never added the annotations he mentions. \par }

\medskip
\leftskip = 0em

We come at length to the letter written by ``Lord Teignmouth, who was long intimately acquainted with this singular man'', and it will be sufficient for our purpose to quote the second of the two long paragraphs which appeared in \textbf{\textit{Obituary}}:

\medskip
\leftskip=2em
{\small\noindent
Mathematics was his favorite pursuit; and perceiving that the science had been cultivated to an extent in Europe far beyond what had been done in Asia, he determined to acquire a knowledge of the European discoveries and improvements; and, with this view, began the study of the English language. He was at this time between forty and fifty; but his success was rapid; and in two years he was not only able to understand any English mathematical work, but to peruse with pleasure the volumes of our best historians and moralists. From the same motive he afterwards studied and acquired the Latin language, though in a less perfect degree; and before his death had made some progress in the acquisition of the Greek dialect. \par}

\medskip
\leftskip = 0em

Campbell winds up \textbf{\textit{Obituary}} by adding a solitary concluding sentence: ``We have nothing to add to this summary of his qualifications and endowment, except our anxious wish, that the whole account may have been rendered sufficiently interesting to reward the perusal of those, who are best able to estimate the merits, and discriminate the peculiarities of his character''.

To fulfil Campbell's anxious wish, some two centuries after he expressed it, is the purpose of this article.

As Lord Teignmouth's letter of appreciation says nothing about Tafazzul's writings, we dip into his \textit{Memoir} (where Tafazzul's name is spelt as Tufuzzool Hossein Khan), and we find there \cite[p.~403]{Teignmouth1843Memoir}: ``His fame as a scholar and a mathematician was established by a Translation of Newton's `Principia' into Persian, and an original Treatise on Fluxions''. We will return to this two-part remark in Section~\ref{sec:FirstTwo} when we come to speak of the first two English translators of the \textit{Principia}.

It is time now to look at the statements made by some other people who knew Tafazzul well, including the two grandees mentioned by Campbell, namely Sir William Jones and Warren Hastings.

\section{Other witnesses. Part 1: Sir William Jones, Reuben Burrow and James Dinwiddie}

We find at least one mention of Tafazzul in a letter (Nr. 520, dated 13 Sept. 1789) written by Sir William Jones (1746--94) to William Steuart. The first sentence of a rather long postscript reads \cite[pp.~838--40]{Jones1970Letters}: ``Give my best compliments to Major Palmer \& tell him that his friend Tafazzul Husain Kh\={a}n is doing wonders in English \& Mathematicks. He is reading Newton with Burrow, \& \textit{means to translate the Principia}  into Arabick''. At the end of the sentence, Cannon (the editor) has added a footnote to announce, as did Campbell in \textbf{\textit{Obituary}}, that the translation did come to fruition: ``William Palmer (1740--1816) was Resident at Lucknow in 1782 and at Sindhia's Court, 1797--8. \ldots  His former Indian colleague \textit{completed} the Arabic translation''. (My italics). 

Jones and Burrow, being active members of the Royal Asiatic Society, knew each other intimately, and we may safely conclude that Burrow (not Burrows, not Barlow) was the name of the person who introduced Tafazzul to the works of English and European mathematicians. Jones expresses, in another letter (Nr. 460, written on 17 June 1787), the hope that ``the ingenious author [Burrow]'' would find the time to prepare a \textit{Dissertation on the Astronomy of the Hindus}. To this remark, Cannon added the following footnote: ``Reuben Burrow (1747--92: \textit{D.N.B.}), mathematician and a loyal Society member, never finished his astronomical treatise. Several of his short papers and lists are in Asiatick Researches (ii).'' 

Campbell and Cannon use, when speaking of Burrow,  the phrase ``never added the annotations'' and ``never finished his astronomical treatise'', as if they are censuring him for contriving the early demise which prevented him from fulfilling his promises and plans!

Reuben Burrow was perhaps the only employee of the Company with a genuine mathematical flair. A short account of his life and expertise is needd to understand what role he played in Tafazzul's mathematical training. According to an obituary \cite{Z-1814Burrow}, Burrow sailed for India in 1782, and the 

\medskip
\leftskip=2em
{\small\noindent
first employment after he arrived at Calcutta was private teaching; this we learn from a paragraph which appeared in one of the English newspapers, stating, that a Cashmirean, one of M. Burrow's pupils who understood English, was translating Newton's Principia into Persian!

\par }

\medskip
\leftskip = 0em

At this point Tafazzul understood English (but not Latin). Burrow, who knew Latin and some French before he set sail for India, seems to have acquired a working knowledge of Persian after his arrival in India (see below). It is conceivable that, after Tafazzul and Burrow became well acquainted, the duo \textit{planned} to translate (into Arabic and/or Persian) some important tracts written by English and French mathematicians. However, Burrow soon found a well paid job in the Company, and was very active in the \textit{Asiatic Society}, which means that he could not have devoted much time to supervise (or collaborate with) Tafazzul, who too was busy with political conjuring and delivering theological lectures. Burrow's unexpected death in 1792 must have been a great blow to Tafazzul. 

The arrival (in September 1794) of James Dinwiddie, a scientific entrepreneur and odd-jobber, passionate advocate of experimental science, and presenter of scientific shows, must have been a godsend to Tafazzul. Dinwiddie's letter (dated 27 Feb 1796) to Joseph Hume confirms the conjecture \cite[p.~134]{Proudfoot2015Dinwiddie}:

\medskip
\leftskip=2em
{\small\noindent
The only good mathematician I have met with, in this country, is a native, the Nabob of Oude's vakeel---his name Tuffoz-ul-Hussien. He is well-known to Mr. Hastings, who sends him out mathematical books. He has translated Newton's Principia into Arabic; also Maclaurin's Fluxions, and the uncouth Emerson's Mechanics. He has been a constant attendant on me since my arrival in Bengal, and is extremely pleased to see the application of theory to practice. Of the latter he had not the least knowledge. 
\par }

\medskip
\leftskip = 0em

Fluxions are mentioned again, but now Tafazzul is said to have merely translated Maclaurin's book, not authored one of his own.

\section{Other witnesses. Part 2: Warren Hastings}
\label{sec:Hastings}
A book critical of the impeachment proceedings against Warren Hastings was published in 1790 \cite{Broome1790Elucidation}. Its author, Ralph Broome (d.~1805), had learnt enough Persian to earn his bread as a translator for the Company, and was ``well informed in Oriental laws, Mohammedan manners, and British transactions in Hindostan'' \cite{Anon1790Monthly}. He was probably the person whom David Anderson called the ``learned Mr. Broome'', Tafazzul's first guide to western mathematics.  At any rate, Ralph Broome, the author of the said book, described an incident that is worth recalling here \cite[p.~viii]{Broome1790Elucidation}. After a chance encounter between Broome and Hastings, they dined together, and during the dinner Hastings 

\medskip
\leftskip=2em
{\small\noindent
expressed a wish that I would call on him at his house in town, where he wanted to show me an Arabic book, which was given to him by a native of India as a mathematical treatise, and supposed to contain problems unknown to Europeans.---I understood that I had been mentioned to him by some of the natives as the likeliest Englishmen [sic] to translate such a work, as it required a knowledge of Arabic and Mathematics, two kinds of learning seldom united in one person residing in India. 

This invitation I did not accept, nor did I ever see him again till long after he was impeached. 
\par }

\medskip
\leftskip = 0em

If Broome was fluent in both Arabic and mathematics, he \textit{was} learned.

In the year 1798, the East India Company appropriated a room for the new building at the India House, to serve as an Oriental Repository, and they invited their servants in India to deposit valuable oriental works in it; on the 18th of February 1800, they appointed Charles Willkins, Esq., to be their Librarian \cite{OrientalRepos1}. On the 23rd of February 1809, Warren Hastings wrote a long note of enquiry to Wilkins, parts of which are reproduced below \cite{OrientalRepos2}: 

\medskip
\leftskip=2em
{\small\noindent
To Charles Wilkins, Esq.\\
Sir,\\
Being desirous of making a sale of all my Persian, Arabic, and Sanscrit, I think it my duty, independently of my interest, to make the first tender of them to the East India Company, for their valuable museum under your charge. \ldots\quad\ldots\quad  Of their value, I have no standard, or other means, for forming an estimate, and wish to submit it to the same authority to which I have referred my first proposal, if this shall be accepted; candidly confessing, that the books, whatever may have been my original purpose in collecting them, are of no use to me now, but in the pecuniary profit which I may derive from the disposal of them.\\
\hspace*{4em} I have the honour to be, \&c.\\ 
\hspace*{6em} Warren Hastings, \&c. 
\par }

\medskip
\leftskip = 0em

A month later, Hastings wrote again:  ``My dear Wilkins, I am going back to the country immediately. Be so kind as to ask the Chairman whether he will consent to take my books for the Company. I cannot transport them back again, and their warehouse hire will be an accumulating charge to me. I have, therefore, made my determination, which depends for its immediate performance on his. This do, my dear friend, obtain for me, and let me know it as soon you are in possession of it.---Yours affectionately, Warren  Hastings;  the 23d of March 1809.''

Wilkins replied with the following valuation: Persian and Arabic books (190 in all) at \pounds 3 a piece and 12,120 leaves of Sanskrit and Hindovi material at  2s. 6d. for eight leaves. Hastings was offered the total sum of \pounds 759 7s. 6d; on the 7th of April 1809, he wrote:

\medskip
\leftskip=2em
{\small\noindent

My dear Wilkins, I thank you for the trouble which you have taken.  I approve of your valuation of my books, and should have approved of it, if it had been less. Yet, I own, I wish that a separate estimate were made of the mathematical books, because I have been told that some of them are curious and uncommon, and two of them [of a set of three (see below)] are beautifully written and drawn, and well selected. They are from the hand of Tofuzzel Hossein Khaun. I will tell you frankly, that I had made up my mind to present them to the Company, if the Chairman made any demur about the purchase.  Of course I leave the disposal of them wholly to your judgment, and final determination. \ldots\\ 
\hspace*{4em} Yours affectionately,\\
\hspace*{6em}  Warren Hastings \&c. \par }

\medskip
\leftskip = 0em

The final determination remained the same as the first, and the two manuscripts of Tafazzul (in his own calligraphic handwriting) fetched Hastings the paltry sum of \pounds 6. Unfortunately, Hastings did not reveal the contents of the two manuscripts, nor whether any of his other ``mathematical books'' were also ``curious and uncommon''. Had an Arabic (or Persian) translation of Newton's \textit{Principia} been among the books he wanted to get off his hands, would Hastings have been just as demure about the pearl in his Indian collection, and Wilkins equally thrifty?

A set of three MSS. (items 743--745 in Otto Loth's catalogue \cite{ArabMS1877-IOLib} of the Arabic books in the India Office Library) fit the description given by Hastings; the catalogue adequately describes the contents of each item, and the cover pages of the online versions \cite{Qatar1249,Qatar0923,Qatar0924} provide some additional, non-trivial details. We need only a condensed description here. Items 743 and 744 are copies of earlier translations (in Arabic or Persian) of the first and second parts (respectively) of a collection, edited by Na\d{s}\={\i}r al-D\={\i}n \d{T}\={u}s\={\i} ({\setarab\novocalize\<na.sIr al-dIn .twsI>}), of ancient Greek treatises; item 745 is a copy of a pre-existing Arabic translation of Books 1--7 of \textit{Conics} of Apollonius of Perga. 

The script, ornamentation and binding of the volumes indicate that all three belong to a set of mathematical tracts. The name(s) of the copyist(s) are not recoreded. Two features of item 745 are not shared by its companion volumes. First, according to my judgment, its scribe is not the same as the person (Tafazzul, according to Hastings) who copied the \d{T}\={u}s\={\i} selections. Second, one finds, between front papers i and ii, a loose leaf of paper on which is inscribed a brief note in Persian, written in the free running scribbling style known as \textit{shikasta}. This is evidently a personal message for someone whom the writer addresses in the top line as ``{\setarab\novocalize\<_hAn .sA.hib mu^sfiq w mahrbAn>}'' (meaning \textit{benevolent and kind kh\={a}n \d{s}\={a}heb}); the message in the next two lines---about deferment of a meeting between the writer and the recipient (line 2) and a prescription for constipation (line 3)---will be of interest only to such as want to know, for academic or personal reasons, the composition of the laxative.

\section{Calcutta School-Book Society digs out three mathematical MSS. of Tafazzul}
\label{sec:3MSS}
In 1817 the Calcutta School-Book Society (CSBS) was formed for answering the demand for printed books of a sufficiently high quality at low prices \cite{Ohdedar1966Growth}. In their second annual report, the Society made an exciting announcement \cite[pp.~17--8]{Calcutta18192ndReport}:

\medskip
\leftskip=1em
{\small\noindent
Three valuable Mathematical MSS. compositions of the celebrated Tufuzz{\it oo}l H{\it oo}syn Khan (the Prime Minister of the late Nuwwab Vuzeer Asif'{\it oo}d Dowlu{\it h}) and the property of his son Tujumm{\it oo}l H{\it oo}syn Khan of Lukhnow, have been kindly lent to this Society in order to their being copied.  One of them exhibits a view of the Copernican System of Astronomy,  the other two are Algebraic treatises. It may be attributed to the enquiries of your Committee that  their existence has come to light, and  \textit{their preservation been secured}.*  [Emphasis added here.]\\\par
\vspace{-1em}
\noindent------------\\ 
\noindent \noindent {\bf FN}: * A brief notice and extract of these works drawn up by Mowluvee Hydur Ullee is intended to form a part of the Persian abstract account of the Society's proceedings and Report for the past year, illustrated by a copper-plate engraving of the Solar system. \par }

\medskip
\leftskip = 0em

Presumably, this Persian document (to be named henceforth as \textbf{\textit{Persian Synopsis}}) was given a meaningful Persian title by its author. A hint as to what this title might have been may be gleaned from a catalogue of the Library of the British Museum \cite{Blumhardt1898Catalog}, where its identification tag is 14117. a. 2 (1). It makes two appearances, once under the name of the author and once under Tafazzul's name (see Figure~\ref{fig:3}). One notices that the Persian titles in the two entries are not identical, that the first has definitely been truncated, and that the second might have been abbreviated. It is curious that the singular form {\setarab\novocalize\<kitAb>} (\textit{kit\={a}b}) has been used here.

\begin{Figure}
 \centering
 \includegraphics[width=1\linewidth]{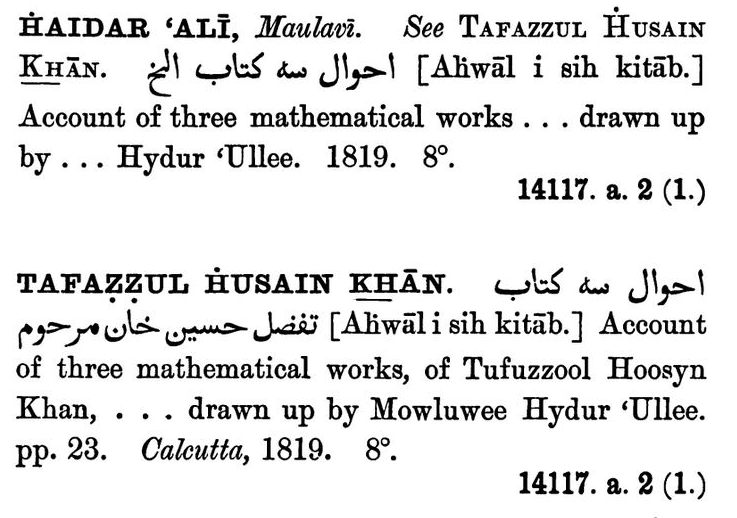}
 \captionof{figure}{Two entries for \textbf{\textit{Persian Synopsis}} in a catalogue of the Library of the British Museum \cite{Blumhardt1898Catalog}. The upper entry occurs in column 108; the lower, in column 330}
\label{fig:3}
\end{Figure}
\medskip
Two other cataloguers list \textbf{\textit{Persian Synopsis}}, but skip the Persian title; Zenker \cite{Zenker1846Bib1079} gives the same English title as that stated by Blumhardt \cite{Blumhardt1898Catalog}, whereas Deloncle \cite{deTassy1879Catalogue588} provides an equivalent French description in which Tafazzul's name is misspelt  as ``Cafazzoul Houssein-Khan''. This 23-page summary (of \textit{three} tracts, written by Tafazzul in Arabic) has now become a rare book. Though I have not been able to find \textbf{\textit{Persian Synopsis}} in the public domain, the effort to find it did dredge up some additional information in the pages of a recent commercial catalogue \cite[p.~43]{BooksOfAsia}, and of the publicity brochure for an auction \cite{DoyleAuction}. The most valuable part of the additional information is an illustration with the legend {\setarab\novocalize\<swlar sistam murattabah .hakIm kwparnikus>} (meaning \textit{the solar system as organized by the philosopher Copernicus}).

We know that Tafazzul's son Tajammul lent three unabridged MSS. to the CSBS for the purpose of being copied. What happened to the copies and to the originals themselves, which were presumably in Tafazzul's own hand, and which must have been returned to Tajammul? Were there any copies prepared within the author's lifetime by professional scribes? None of these questions I am able to answer with certainty, but I can report reading the following remark in the same report where their recovery had been announced with so much joy \cite[p.~43]{Calcutta18192ndReport}:

\medskip
\leftskip=2em
{\small\noindent
The MSS. some months ago obtained from Lucknow, the compositions of the famous Tufuzz{\it oo}l H{\it oo}syn Khan, and in a fair way of becoming food for \textit{worms}, not \textit{students}, furnish a case in point to shew the truth of Mr. Robinson's observations, that our interposition may save valuable performances from perishing. \par }

\medskip
\leftskip = 0em

The Society's sense of achievement---that, thanks to the efforts of their Committee, the ``preservation [of the three Tafazzul manuscripts had] been secured''---seems to have been premature (or dashed, if it was just an earnest hope). The Committee members might have concluded that worms had caused enough damage to prevent a mathematically untrained scribe from making useful copies. The worms, it seems, were allowed to do their work for a few more years (see below).

\section{John Tytler connects with Indian mathematicians}

John Tytler (1787--1837) arrived in India in 1813, and was assigned in the beginning of 1814 to the civil station of Patna. There he met
D\={\i}w\={a}n Kanh J\={\i}, an Indian scholar who had prepared a compendium (written in Persian) of whatever mathematics he could learn from Indian and foreign sources; fortunately, he included the contents of \textbf{\textit{Algebra}}  and \textbf{\textit{Algebra-in-Geometry}} in two sections of the compendium. Referring to Tafazzul as ``the late Tafazzul Husain Khan'', the D\={\i}w\={a}n informs the reader that Tafazzul prepared his material by translating English books into Arabic. The first of these sections (pp.~546--579) covers algebra and the second (pp.~579--624) discusses the solution of geometrical problems by means of algebraic analysis. The printed version of the compendium, which provides the English equivalent of all important terms (written in Persian characters),  uses the term ``geometric algebra'' for what may also be named ``algebra-assisted geometry''. The compendium is commonly dubbed \textit{Khazanat al-Ilm} (also \textit{Khazanat-ul Ilm}), though the full title is a little longer and far more informative \cite{Khazana1837KanhJi}.  The contents of the compendium inspired Tytler to publish two articles \cite{Tytler1820AsiaticRes,Tytler1832AsiaticRes}, the first of which concluded with the following words \cite{Tytler1820AsiaticRes}:

\medskip
\leftskip=2em
{\small
It is but justice that I should add, that my first knowledge of this rule was obtained from the \textit{Khazanut-ul Ilm}, which is a complete system of Arithmetic, Algebra, and Geometry, as far as known to the \textit{Arabians} and \textit{Hindus}, composed in the present day by Khan Jee, a most intelligent inhabitant of \textit{Patna}. On my requesting to know from what original authors the rule was taken, this gentleman was kind enough to favour me with the above extract. No more I think is required to demonstrate, that his own work highly deserves translation and publication.\par } 

\smallskip
\leftskip=0em

The publication of Kanh J\={\i}'s \textit{Khazanut-ul Ilm} became a printing ordeal, which is described already in the middle of the title page. The 
description is divided into three centred blocks of text in the format shown below:

\begin{center}
{\small
Adopted for publication, by the General Committee of Public Instruction, for the general use of the Persian Colleges under their control, and printed up to the 492nd page under the supervision of Dr. J. Tytler.\\[4truept]

Suspended by order of Government, and transferred with other unfinished Oriental works to the Asiatic Society, in March 1835,\\[4truept]

and completed at the Society's expence, under the gratuitous supervision of Maulav\={\i} Mans\={u}r Ahmed Bardaw\={a}n\={\i}, one of the teachers at the College Haji Mohsin, Hoogly, September, 1837.
}
\end{center}
Failing health compelled John Tytler to discontinue his involvement with the book; he decided to leave India, and arrived in England in May 1835. Despite being ``broken in health, depressed in spirits, and impaired in fortune'' \cite{TytlerObituary1837}, Tytler spent a great deal of energy on promoting the work of ``a Maulavi, by name Gholaum Hosain'', who had come to see Tytler (shortly before his departure) in the hope that the latter would recommend his manuscript to the Government Education Committee. Tytler obliged by endorsing the book enthusiastically in a long and characteristically thorough article, published posthumously \cite{Tytler1837PersianWork}.

Whether or not Ghul\={a}m Husain received any financial assistance from the Education Committee is not known (at least to me), but we know from an 1838 report on the state of education in Bengal that the book \textit{was} published \cite{Adam1838Third}: 

\smallskip
\leftskip=2em
{\small
Maulavi Gholam Hossein, dwelling at Sahebgunge in the thana of that name, has written in Persian a compilation called \textit{Jam-i-Bahadur Khani}, from various Arabic works on arithmetic, geometry, astronomy, and the natural sciences with additions of his own. This work has been printed and contains 720 pages. He is now engaged in the preparation of astronomical tables to be entitled \textit{Zij Bahadur Khani}. The names of both works are intended as a compliment to his patron Bahadur Khan, one of the sons of Mitrajit Singh, the Raja of Tikari. \par}

\medskip
\leftskip=0em

Passionate about the publication of scientific books in Arabic, Tytler wrote as well as translated several pieces himself. To quote from his obituary \cite{TytlerObituary1837}:

\medskip
\leftskip=2em
{\small\noindent
The interest first imbibed by Mr. Tytler at Patna from the example and aid of his friend Khan Ji, in these arduous and abstruse enquiries, continued unimpaired to the last: and, after his return to England, he communicated to the Royal Asiatic Society an analysis of a work analogous to the \textit{Khazanat al Ilm}, named the \textit{Jamia Bahadar Khani}, by Maulavi Gholam Hosein, a scientific native in the service of Buhadar Khan, the son of the Raja of Tikari, another of Mr. Tytler's Patna friends, and a patron of mathematical learning. Mr. Tytler also prepared for the Ashmolean Society of Oxford, an account of an Arabic version of the Conic Sections of Apollonius, and of other mathematical works originally written in Europe, of which he had brought home a manuscript copy. \par } 

\medskip
\leftskip=0em

So much space has been devoted to Tytler in an article about Tafazzul, because if a presentable Arabic translation of the \textit{Principia} had existed, Tytler would probably have been informed by his Indian ``men of mathematics''. I have not come across a reference to any of Tafazzul's putative translations in one of Tytler's articles. However, the rumour must have been floating around, as may be judged from the concluding remarks, in a few pages devoted to book reviews. After spending one paragraph on the Indian and Arabian part of \textit{Khazanat al-Ilm}, the reviewer states \cite{Anon1833JAsiaSocBengal}:

\medskip
\leftskip=2em
{\small
The European part of the Khizanut-ool-Ilm consists of two sections: first, a complete translation by the Dewan of Bonnycastle's Algebra; secondly, an extract consisting of a collection of Geometrical Problems from the papers of the celebrated {\sc Tufuzzool Hosain Khaun} of Delhi. This person during his life, was considered we believe, the best Mohammadan mathematician in India, and he appears to have employed his time [all his time?] in translating European mathematical works into \textit{Arabic}; after his death, which took place some years ago, Government, we are told, made strong efforts to obtain his MSS, but in consequence of legal disputes between his relations these were unsuccessful, and the fate of the papers is probably not known. It is much to be wished that they could be procured.\par } 

\medskip
\leftskip=0em

Tafazzul did spend, we happen to know, a few years in Delhi but many more in Lucknow, and---not by the bye---the second section of \textit{Khazanat al-Ilm} is not (as already stated above) quite ``an extract of Geometrical Problems from the papers of'' Tafazzul. The rest of the statement may also have been vitiated by some imprecision or involuntary distortions, but perhaps not enough to have squeezed all truth out of the words. If that is granted, one may go on to conclude that, some thirty years after his demise, the unravaged portions of the mathematical papers of Tafazzul---which might have included some presentable translations---were still in the hands of his heirs. We also have to assume that the reviewer was not speaking of the three short tracts that had been loaned by Tafazzul's son to CSBS.

\section{Tafazzul's mathematical works: separating rumour from fact}

By now we have heard so many claims on behalf of Tafazzul that it is not easy to recall who said what. Did Tafazzul translate the \textit{Principia} into Arabic or Persian? Did he use the original Latin text of Newton or an English translation? Did he also write an original treatise on fluxions, or did he merely translate Maclaurin's tract? Did he translate other books as well? Emerson's Mechanics? A book on conic sections by de l'Hospital and another on the same topic by Robert Simson? Did he also translate Burrow's English version of a book of Apollonius? Did he translate Simpson's book on algebra, or that written for him by Reuben Burrow, or both?

It will be well to recall, before attempting to answer the above questions, what Allport and Postman wrote about rumour \cite[p.~43]{Allport1948Psychology}: ``To be sure, in rumour there is often some residual particle of news, a ``kernel of truth,'' but in the course of transmission it has become so overlaid with fanciful elaboration that it is no longer separable or detectable. In the rumoured story it is almost always impossible to tell precisely what the underlying facts are, or indeed whether there are any at all.'' 

Let us recall SMA's words in \textbf{\textit{UrduBio}} (see Figure~\ref{fig:2} or the English rendering which follows it): ``He wrote two algebraic treatises, one on algebraic solutions [i.e. \textbf{\textit{Algebra}}] and the other on algebro-geometric solutions [i.e. \textbf{\textit{Algebra-in-Geometry}}]. (Indeed, I have seen some pages of the second treatise, published in Calcutta).'' The first sentence is correct, but SMA is just echoing Sh\={u}shtar\={\i} here, for he has not seen either of these tracts; the few pages of \textbf{\textit{Algebra-in-Geometry}} published in Calcutta that crossed his eyes must have been in \textbf{\textit{Persian Synopsis}}, or in \textit{Khazanut-ul Ilm} \cite{Khazana1837KanhJi}, which were both written in Persian, not Arabic.

On the basis of what has been presented above, we should readily agree that \textbf{\textit{Algebra}}, \textbf{\textit{Algebra-in-Geometry}}  and \textbf{\textit{Copernican Astronomy}} form the kernel of truth in the reports concerning Tafazzul's prolific authorship of mathematical works, and these are precisely the three tracts Beale chose to mention in \textbf{\textit{Chronograms}}.

Had there been no trace of  \textit{Khazanut-ul Ilm}, whose author was adept at mathematics, besides being a contemporary of Tafazzul, we would not have been half as confident about the existence of \textbf{\textit{Algebra}} and \textbf{\textit{Algebra-in-Geometry}}. Before proceeding to look at claims concerning translations of specific mathematical treatises, it will be helpful to recall a similar case from the distant past.

Aristarchus of Samos has come to be recognised as the first known proponent of a heliocentric model for our solar system, although the proposal is not to be found in his only extant manuscript \cite{Heath1913aristarchus,Evans2021EB}. The attribution is based on the authority of Archimedes, a virtuoso astronomer himself; it need hardly be said that, though we must be prepared to listen to less eminent witnesses, we should not take, when speaking about a matter of such importance, the word of someone with a questionable competence. The absence of a manuscript that is claimed to have been written in the near or distant past need not be an impediment to its being attributed to a given author, but the requirement about the credentials of those who vouch for its existence at some point in time, for its author's identity and it contents, cannot be relaxed.

In Tafazzul's case, we should not be expected to take \textit{anyone}'s word about his written works, unless the testimony comes from a person who has seen Tafazzul's translation of the \textit{Principia}, has read one or two chapters, and has the requisite competence, which amounts in this case to possessing the capacity to comprehend written Arabic and/or Persian \textit{and} knowing enough mathematics to be able to judge the fidelity of the translation. Among Tafazzul's contemporaries, the only one who does not fall through the sieve of competence is Reuben Burrow, and he did not go beyond saying that Tafazzul ``continues translating the Principia of Newton''; we may even give Sir William Jones the benefit of the doubt, but he too went no further than announcing (only once) that Tafazzul, ``is doing wonders in English \& Mathematicks'',  and that he ``is reading Newton with Burrow, \& means to translate the Principia into Arabick''. Among those who arrived after Tafazzul's decease, John Tytler was uniquely qualified to judge the issue, but (to the best of my knowledge) he did not mention coming across any work of translation by Tafazzul.

Was Tafazzul an impostor, then? A considerable amount of preliminary discussion about the very task of translating a mathematical treatise is needed before we attempt to answer this question.

\section{Different types of translations}

Let us refer to the opening remark of a passage (quoted above) from \textbf{\textit{UrduBio}}, in which SMA extolls Taffazul's mastery of foreign languages: ``He also had complete command over English, Latin and Greek, as may be judged from the works written, edited and translated by him.'' We will be able to judge only when we find a manuscript or two. Apart from this obvious objection, it is crucial to emphasize that, since mathematics itself is a language, a translator who has nearly the same grasp of the subject as the author (of the work to be translated) need not be an expert in the source language.

For a striking illustration of the truth of the above statement, we need only return to Burrow's account of the translation, from Arabic to Latin, of \textit{De sectione rationis}, and supplement it with the nitty gritty of this particular translation project. The original Greek version was lost (but not before it had been translated into Arabic). The Bodelian library had acquired an Arabic manuscript through the efforts of Edward Bernard, who began, but did not complete, a translation of it, and the task eventually fell to Edmond Halley, who succeeded, \textit{though he knew no Arabic}, in producing a translation that has been much admired! In the preface to his Latin translation \cite{1706ApollonSectione}, Halley stated the curious circumstances under which the translation was performed. Though an English rendering of the relevant passage is available \cite[p.~221]{Molland1994Limited}, our purpose will be better served by quoting a longer excerpt from a memoir that was published along with Halley's correspondence 
and papers \cite{MemoirHalley1937}:   

\medskip
\leftskip=2em
{\small 
In 1706 he publisht Apollonius's book, \textit{De sectione Rationis}, by him translated, or rather decypher'd, from an Arabic Manuscript, in the Bodleian Library; for he did not, at that time, understand the Arabic Tongue, but only translated the whole by the assistance of a very few pages of it already translated by Dr. Barnard, which he made use of, as a Key to the rest; and this he did with such success, through his being so great a Master of the Subject, that I remember the Learned Dr. Sykes, (our Hebrew Professor at Cambridge, and the greatest Orientalist of his time, when I was at that University,) told me, that Mr. Halley talking with him upon the subject, shew'd him two or 3 passages which wanted Emmendation, telling him what the Author said, and what he shou'd have said, and which Dr. Sykes found he might with great ease be made to say, by small corrections, he was by this means enabled to make in the Text. Thus, I remember, Dr. Sykes expresst himself, Mr. Halley made Emendations to the Text of an Author, he could not so much as read the language of.\par}

\medskip
\leftskip=0em

When it comes to the translation of Newton's \textit{Principia}, the magnitude of the task presented, in the first few decades after its publication, a far bigger challenge, because the novelty of the approach put the translator at an enormous disadvantage. The person who translated the \textit{Principia} into French was the Marquise du Ch\^{a}telet, who knew Latin and English, and was no mean mathematician herself \cite{HuffmanAMMS,Zinsser2001French}. Like Tafazzul, she led a busy life, but she spent all her spare time on converting Newton's text into a version that would be intelligible to the mathematicians of Europe, most of whom were more familiar with French than with English, and were in tune with the new concepts, nomenclature and notation pioneered by Leibniz. \'{E}milie du Ch\^{a}telet saw no point in slavishly translating the \textit{Principia} into French; she  decided instead to re-dress Newton's arguments in the garb of Continental calculus, and was thereby able to consult two eminent French mathematicians, who were part of her circle of friends, had embraced and digested Newton's ideas, but rejected the cumbersome mathematical apparatus he used for preparing the \textit{Principia}. As for Newton's text itself, Zinsser has quoted from du Ch\^{a}telet's letter to Daniel Bernoulli: ``m. Neuton's Latin is one of the difficulties'' \cite{Zinsser2001French}. 

For Tafazzul, whose knowledge of Latin could not have been other than superficial, and an understanding of Newtonian physics hardly better, making an Arabic translation that adhered closely to the Latin version of the \textit{Principia} must have been a tough nut to crack, even with a helping hand from Burrow. 

Both Halley and Mme du Ch\^{a}telet wanted to, and succeeded in, preparing presentable translations. There also arise occasions when a reader wants to prepare a translation only for their own benefit. One learns, after observing capable collaborators who have a totally inadequate knowledge of English, that many of them compensate for their linguistic deficiency by ``translating'' into English the key papers in their discipline by writing manually the meanings (in the language of their choice) of all unfamiliar English words and phrases, placing the dictionary equivalents, whenever possible, directly above the relevant part of the printed English text. With perseverance they manage to garner an informal, syntactically dishevelled, set of word strings that answers their needs even if the whole falls far short of a presentable translation. When the documents happen to be photocopies of an original, the translator is free to use a pencil or write with a pen (using, when helpful, inks of different colour). If copies are not easy to make, most people would use a pencil, and this, Hutton tells us \cite[p.~64]{Hutton1815PhilDict}, is how Burrow prepared translations for his own use:

\medskip
\leftskip=2em
{\small 
The late Mr. Reuben Burrow collected, in India, many oriental manuscripts on the mathematical sciences, both in the Sanscrit and the Persian languages, the latter being translations only of the former: most of these he bequeathed by will to one of his sons there \ldots . But one or two of these Burrow left to his friend Mr. Dalby, 
\ldots being the Persian translations of the Bija Ganita and Lilawati, with an attempt at an English translation of some parts of them by Mr. Burrow; but these attempts being mostly interlineations written with a black-lead pencil, are in danger of being obliterated. 
\par}

\medskip
\leftskip=0em

Among  the people who spoke of Tafazzul's translations, only Burrow said ``we shall soon begin to print it here in Arabic'', and one learns that his ``notes and explanations'' were meant to be an integral part of the \textit{printed} and presentable translation of the \textit{Principia}. Burrow's premature death made that hard nut harder still---well-nigh impossible for Tafazzul to crack on his own.

\section{Who remembers the first two English translators of \textit{Principia}?}
\label{sec:FirstTwo}

Of the first two English translations of Newton's  \textit{Principia}, only that prepared by Andrew Motte and published in 1729 is widely known. According to Bernard Cohen \cite{Cohen1963TwoTranslators}, ``a second and independent translation appears to have been made at the same time by Henry Pemberton, Newton's collaborator and disciple''. Pemberton knew some mathematics, but no one would have called him a mathematician; interesting details about his translation, which did not prove to be remunerative, may be found in Ref.~\cite{Cohen1963TwoTranslators}. As for Motte, Cohen writes: ``We know next to nothing about Andrew Motte. I have been unable to find any information concerning his life, e. g., his date of birth, his education, or his literary and scientific career. Nor have I found out when he undertook to translate the Principia, or under what circumstances''. The only trace he has left as a man of letters is the following line: ``Mr Andrew Motte,  Author of  \textit{The Laws of Morion}, and several other Tracts in the Mathematicks.''

Let us return now to the remark from the \textit{Memoir} of Lord Teignmouth: ``His fame as a scholar and a mathematician was established by a Translation of Newton's `Principia' into Persian, and an original Treatise on Fluxions'', and just suppose that the sentence had finished after the word ``Persian''. I wonder if Lord Teignmouth also realized that recognition as a mathematician requires more than the ability to translate a mathematical tract, however magnificent that tract may be, and felt the necessity of conferring extra glory on his friend Tafazzul by crediting him with an original work on fluxions.

\section{Drawing the threads together}
\label{sec:threads}
It has been argued above that many Company employees made assertions (about the nature and extent of Tafazzul's mathematical accomplishments) that were  unsubstantiated and---even if the rumour raisers, Tafazzul included, did not think so---highly implausible. As is typical of situations where rumour thrives, the Company servants transmitted information that they had not bothered to, or failed to, contextualize when they first received it. Was it this author or that? this book of a particular author or that one? translated from Latin or from English? into Arabic or into Persian? translated a book on some abstruse topic or wrote one himself?

Rumours are topical, therefore short-lived. Any rumour that lives long enough turns into a myth. It remains to be shown that the rumour (about Tafazzul's  translation undertakings), which lay dormant for more than a century, has been gaining currency among contemporary scholars, and is now turning into a myth. Challenging a potential myth, rather than delving into the undeclared motives, either of the originators of the rumour or of its more recent perpetuators, has been my aim here.

\section{Revival of the rumour}

The list of authors who have accepted one or more of the claims asserted without proof in \textbf{\textit{Tuhfa}} and \textbf{\textit{Obituary}}, or by other friends of Tafazzul, and used it to buttress one or another point of view is rather large; otherwise the use of the term ``rumour'' would not have been warranted. Some of the works which have propagated the rumour, and the supporting references cited therein, are displayed in Table~\ref{tab:diffusers}. 

\end{multicols}

\begin{center}
\renewcommand{\arraystretch}{1.2}
\begin{longtable}{|r|l|l|l|l|l|}
\caption{A selection of works whose authors have participated in the diffusion of the rumour concerning Tafazzul's translation endeavours.} \label{tab:diffusers} \\
\hline \multicolumn{1}{|r|}{\textbf{Nr.}} & \multicolumn{1}{c|}{\textbf{Year}} & \multicolumn{1}{c|}{\textbf{Author}} & \multicolumn{1}{c|}{\textbf{Type}}&  \multicolumn{1}{c|}{\textbf{Ref}} & \multicolumn{1}{c|}{\textbf{Principal Sources}} \\ \hline 
\endfirsthead
\multicolumn{6}{c}%
{{\bfseries \tablename\ \thetable{} -- continued from previous page}} \\
\hline \multicolumn{1}{|r|}{\textbf{Nr.}} & \multicolumn{1}{c|}{\textbf{Year}} & \multicolumn{1}{c|}{\textbf{Author}} & \multicolumn{1}{c|}{\textbf{Type}}&  \multicolumn{1}{c|}{\textbf{Ref}} & \multicolumn{1}{c|}{\textbf{Principal Sources}} \\ \hline 
\endhead

\hline \multicolumn{6}{|r|}{{Continued on next page}} \\ \hline
\endfoot

\hline \hline
\endlastfoot
1	&1880 		&Mu\d{h}ammad \d{H}usain {\=A}z{\=a}d	&Book				&\cite{Azad1907AbeHayat} 	&										\\    
2	&1940 		&A. Yusuf Ali					&Book				&\cite{YusufAli1940Cultural} 							&\cite{Campbell1804AARegister}\\  
3	&1943		&Purnendu Basu				&Book				&\cite{Basu1943Oudh}									&\cite{Campbell1804AARegister}\\   
4	&1945		&R. V. Parulekar				&Book				&\cite{Bombay1945Survey}							&\\   
5	&1963		&Mulk Raj Anand				&Book				&\cite{Anand1963ContemCivil}						&\\
6	&1986		&S. Athar Abbas Rizvi		&Book				&\cite{Rizvi1986Socio}									&\cite{Shushtari1847Tuhfa}\\
7	&1993		&Gulfishan Khan				&PhD Thesis		&\cite{Gulfishan1993Perception}						&\cite{Shushtari1847Tuhfa,Campbell1804AARegister}\\	
8	&1995		&Fran\c{c}ois Charette		&Master's Thesis	&\cite{Charette1995Orientalisme}					&\cite{Jones1970Letters}\\
9	&1996		&M. Tavakoli-Targhi			&Article				&\cite{Tavakoli1996Orient}							&\cite{Shushtari1847Tuhfa,Jones1970Letters}\\
10	&1999		&C. A. Bayly					&Book				&\cite{Bayly1999Empire}								&\cite{Rizvi1986Socio}	\\
11	&1997		&Francis Robinson			&Article				&\cite{Robinson1997OttoSafaMugh}				&\\
12	&2000		&Gail Minault					&Chapter			&\cite{Minault2000Qiran}								&\cite{Campbell1804AARegister}\\
13	&2001		&Francis Robinson			&Book				&\cite{Robinson2001Ulama}							&\\
14	&2001		&M. Tavakoli-Targhi			&Book				&\cite{Tavakoli2001Refashioning}					&\cite{Shushtari1847Tuhfa,Campbell1804AARegister}\\
15	&2002		&Tariq Rahman				&Book				&\cite{Rahman2002Language}						&\cite{Gulfishan1993Perception}\\
16	&2003		&William Dalrymple			&Book				&\cite{Dalrymple2003White}							&\cite{Shushtari1847Tuhfa}\\
17	&2003		&M. K. Chancey				&PhD Thesis		&\cite{Chancey2003EIC}								&\cite{KhanSMA1908TafazLife}\\
18	&2003		&Iqbal Ghani Khan			&Article				&\cite{IGKhan2003IndJHS}							&\cite{Shushtari1847Tuhfa}\\
19	&2004		&M. Tavakoli-Targhi			&Chapter			&\cite{Tavakoli2004Homeless}						&\cite{Campbell1804AARegister}\\
20&2008			&Kapil Raj						&Article				&\cite{Raj2008Revue}									&\cite{Shushtari1847Tuhfa,KhanSMA1908TafazLife}\\
21&2009			&Simon Schaffer				&Chapter			&\cite{Schaffer2009Astro}	  							&\cite{Campbell1804AARegister,Shushtari1847Tuhfa,KhanSMA1908TafazLife,Rizvi1986Socio}\\
22&2009			&Kapil Raj						&Chapter			&\cite{Raj2009Calcutta}	  							&\cite{Shushtari1847Tuhfa,KhanSMA1908TafazLife}\\
23&2011			&Kapil Raj						&Article				&\cite{Raj2011Calcutta18C}	  						&\cite{Shushtari1847Tuhfa,KhanSMA1908TafazLife}\\
24&2011			&M. Tavakoli-Targhi			&Chapter			&\cite{Tavakoli2011Pollock}							&\cite{Shushtari1847Tuhfa,Campbell1804AARegister}\\
25&2012			&Partha Chatterjee			&Book				&\cite{Chatterjee2012Black}							&\cite{Gulfishan1993Perception}\\  		
26&2014			&Joydeep Sen					&Book				&\cite{Sen2014Astronomy}							&\cite{Schaffer2009Astro}\\
27&2016			&Kaveh Yazdani				&Article				&\cite{Yazdani2016Persianate}						&\cite{Shushtari1847Tuhfa,Anand1963ContemCivil,Rizvi1986Socio,Gulfishan1993Perception,Schaffer2009Astro}\\
28&2017			&Kaveh Yazdani				&Book				&\cite{Yazdani2017India}								&\cite{Gulfishan1993Perception,Robinson1997OttoSafaMugh,Shushtari1847Tuhfa}\\
29&2018			&Joshua Ehrlich				&PhD Thesis		&\cite{Ehlrich2018EIC}									&\cite{Campbell1804AARegister,Robinson2001Ulama,Schaffer2009Astro}\\
30&2018			&Michael Bergunder			&Article				&\cite{Bergunder2018IndiaAstro}					&\cite{Campbell1804AARegister,Gulfishan1993Perception,Rizvi1986Socio,Schaffer2009Astro}\\
31&2021			&Pradip Baksi					&Book				&\cite{Baksi2021Marx}									&\cite{Campbell1804AARegister}\\
\end{longtable}
\end{center}

\begin{multicols}{2}

\vspace{-32truept}
A glance at the last column of Table~\ref{tab:diffusers} will show that Refs.~\cite{Shushtari1847Tuhfa} and \cite{Campbell1804AARegister} (\textbf{\textit{Tuhfa}}  and  \textbf{\textit{Obituary}}, respectively) are the most frequently cited sources, which stands to reason, since these are truly the fons et origo of the rumour. A few additional comments appear to be in order, and these are listed below. The numbers in the list correspond with the row numbers in Table~\ref{tab:diffusers}. 

\begin{enumerate}[label={\arabic*.}, start=1, leftmargin=*]
\item 
Tafazzul's name appears prominently in Mu\d{h}ammad \d{H}usain \={A}z\={a}d's legendary \textit{\={A}b\'{e} \d{h}ay\={a}t} ({\seturdu\novocalize\<Abe .hayAt>}),  first published in 1880 \cite{Azad1907AbeHayat}, and translated into English in 2001 \cite{Azad2001AbeHayat}. This book is cited by SMA as item 8 in the list on p.~7 of the biography \cite{KhanSMA1915TafazBio}. \={A}z\={a}d's passage describing Tafazzul (p.~255) has a footnote with three sentences (in the English version), the second of which reads: ``He had learned the English and Latin languages too; he had translated the \textit{Differential}, and so on, of Newton Sahib into Persian''. \={A}z\={a}d was evidently unaware that Newton Sahib abhorred the \textit{differential} and its creator, and it was wise of \={A}z\={a}d  Sahib to use the escape hatch  ``and so on'' (translation of {\seturdu\novocalize\<wa.gyraH>}), which prevented him from committing further blunders.
\item  Of all the authors listed in Table~\ref{tab:diffusers}, Yusuf Ali is the most cautious, for he wrote \cite[p.~74]{YusufAli1940Cultural}: ``Tafazzul Husain Khan, the Vakil of Nawab Asaf-ud-Daula at Calcutta, about 1788--92, was \textit{engaged in translating} Sir Isaac Newton's \textit{Principia} from Latin into Arabic (or was it Persian?). He also \textit{attempted to translate} books on Algebra, Mechanics, Conic Sections, and Logarithms. He knew many languages, including Greek.  He died in 1800, and a notice of him appeared in the Asiatic Register (Vol. V, 1803; Characters, p. 7).''  (Apart from the book title \textit{Principia}, all italics are mine.)\\
\hspace*{6truept} With no reasons given by Yusuf Ali for his scepticism, the tentativeness of his understatements is more likely to be lost than not.
\setcounter{enumi}{4}
\item  M. R. Anand cites no one for the statement, but his phraseology (``attempted to translate'') suggests that he was influenced by Yusuf Ali.
\setcounter{enumi}{10}
\item[] \hspace*{\dimexpr-\labelindent-\labelwidth-\labelsep+\itemindent}%
11. \& 13. 
Robinson states the following and feels no need to substantiate the statement \cite{Robinson1997OttoSafaMugh,Robinson2001Ulama}: ``Indeed, in the late eighteenth and early nineteenth centuries Lucknow was a major intellectual centre training scholars who took pleasure in engaging with European science such as the polymath Tafazzul Husain, who translated Newton's Principia into Arabic, and \ldots ''.
\setcounter{enumi}{11}
\item Minault states \cite{Minault2000Qiran} in footnote 31: ``In fact, Newton's \textit{Principia} and other works of European mathematics and astronomy that supported the heliocentric view of the universe had already been translated into Arabic and Persian in India in the late 18th century by Maulawi Tafazzul Husain Khan of Lucknow. He is mentioned in the \textit{Asiatick Annual Register of 1803}, pp.~1--8; and in Abu Talib Khan, \textit{Ma\textrevglotstop\={a}thir-e T\={a}lib\={\i}} (personal communication between the author and Md. Tavakoli-Targhi). Cf. Tavakoli-Targhi, ``Orientalism's Genesis Amnesia" and \ldots .''\\[1em]
Two other points must also be made here:--\par
\hspace*{6truept} (a) Anyone who, when alive, was a pillar of the Awadh court, and commanded, when dead, sixteen columns in the \textit{Asiatick Annual Register} couldn't have been a mere Maulawi. However,  Minault is in very good company: even SMA, the descendant who wrote Tafazzul's biography, used this honorific in the title.
Tafazzul is indeed the subject of \textbf{\textit{Obituary}}, but he is credited there with only an Arabic translation of Newton's \textit{Principia}; true,  Ab\={u} \d{T}\={a}lib Kh\={a}n also wrote about Tafazzul, but in a different book, and for a different role, namely a key player in the political affairs of Awadh after Asaf ud-Daula became the Naw\={a}b. \par
\hspace*{6truept} (b) After Minault (2000) has cited \textbf{\textit{Obituary}}, her mention of Tavakoli-Targhi's 1998 article ``Orientalism's Genesis Amnesia" does not add further weight to the translation claim, because the latter author also cites \textbf{\textit{Obituary}} for support. The rumour is continued by Schaffer \cite{Schaffer2009Astro} when he cites (in 2009) the above footnote in a footnote of his own (Nr. 32): ``Gail Minault \ldots cites Tafazzul's translation of Newton in connexion with Muslim students' demand for astronomy teaching in the Urdu curriculum at Delhi College in the 1840s.''
\setcounter{enumi}{15}
\item
William Dalrymple's presence in Table~\ref{tab:diffusers} is easily explained. Sh\={u}shtar\={\i} was related to one of the protagonists in \textit{White Mughals}, wherein Tafazzul is accorded---because he is greatly admired by Sh\={u}shtar\={\i}---a \textit{very} long footnote on p.~271. That Dalrymple is not intimate with Tafazzul may be deduced from the fact that when this dignitary is first presented to the readers (line 7), his full name is declared to be ``Abu Talib Tafazul''!\par
\hspace{6truept} Whatever the cause, Schaffer quotes Sh\={u}shtar\={\i} through \textit{White Mughals}. His indirect quote ends on the following remark \cite[p.~59]{Schaffer2009Astro}: ``Latin, \ldots the learned tongue of the Europeans in which they write their scholarly books, and which has the same position among them as Arabic among non-Arab Muslims.'' Sh\={u}shtar\={\i} may have thought so, but the analogy, useful though it is, breaks down when it is stretched as far eastward as India (see next item for a fuller discussion). 
\setcounter{enumi}{26}
\item[] \hspace*{\dimexpr-\labelindent-\labelwidth-\labelsep+\itemindent}%
27.--28.
Since acceptance of one or both of the primeval sources (\textbf{\textit{Tuhfa}}  and  \textbf{\textit{Obituary}}) is a trait common to nearly all the disseminators, it is perhaps unfair to choose Yazdani as a representative. However, his comments on other sources provide a more telling indication of the danger of suspending one's judgment. In a long footnote full of bibliographic data, Yazdani  mentions that Anand \cite{Anand1963ContemCivil}, Bayly \cite{Bayly1999Empire}, and Rizvi \cite{Rizvi1986Socio} averred that Tafazzul translated the \textit{Principia} into Persian. This, Yazdani argues, ``seems to be unlikely since even in the Persianate world scientific texts were usually written in Arabic'', after which he goes on to cite some other authors who have stressed the singular status of Arabic in the Muslim world. To clinch the argument by removing any lingering doubts, Yazdani refers to Schaffer \cite{Schaffer2009Astro}, because he (Schaffer) ``even provides archival evidence from British contemporaries that Tafazzul translated Newton and other scientific works into Arabic''. Schaffer has indeed taken great pains to collect the comments of Tafazzul’s contemporaries, but their words, being mere assertions, do not satisfy the criterion of secure evidence; what is needed, \textit{at the very least}, is a written assessment of Tafazzul’s translation of the \textit{Principia}. One’s faith in the offhand comments of Tafazzul’s contemporaries is shattered by their lack of unanimity. Recall that Schaffer also refers to ``the collection of primary sources'' in (what has been named here as) \textbf{\textit{Leaflet}}, and one finds, on p.~1 of Part II, the remark ``His fame as a scholar and a mathematician was established by a Translation of Newton's ``Principia" into Persian''.\par
\hspace*{6truept} 
Yazdani has assumed, as did Sh\={u}shtar\={\i} some two centuries ago, that what was true for the Persianate world was true also for India-under-the-Muslim-rule, an assumption that overlooks the fact that the Persianate world was predominantly peopled by followers of Islam, whereas Muslims of India were vastly outnumbered by Hindus; indeed, an examination of the historical development of literature in India does not support the assumption. Both Muslim and Hindu authors adopted Persian as the medium for secular topics; Arabic or Sanskrit (depending on their religion), for religious texts. We have already seen two examples: D\={\i}w\={a}n Kanh J\={\i} (a Hindu) and Maulav\={\i} Ghul\={a}m Husain (a Muslim) both wrote their mathematical texts in Persian. Ironically, it was Tytler who championed, rather quixotically, the cause of Arabic as the ideal medium for disseminating scientific knowledge among Indian Muslims, and prided himself for being their benefactor.\par 
\hspace*{6truept} 
There is no need to build a cogent case by marshalling afresh the pertinent details, for Sherwani has already dealt with the issue capably. He begins the discussion with the apt remark \cite[pp.~81--96]{Sherwani1969Cultural}: ``It is strange that in spite of the virtual hegemony of Muslim rulers in a large part of India during the period under review [ca. 1200--1760], the output of Arabic literature produced in the country was comparatively meagre and was mostly confined to religious topics.'' Also, one should not overlook the fact that a scholar like \={A}z\={a}d did not find it odd that Tafazzul chose to translate ``Newton's \textit{Differential} and so on'' into Persian.
\hspace*{12truept}

\end{enumerate}

\section{Concluding remarks}

That Tafazzul---with only a cursory knowledge of Latin, deprived of able mathematicians' company, and disinclined to turn his back on competing, non-scientific scholarly commitments---managed to complete, during the nine years or so that elapsed between Burrow's death and his own, a presentable translation of not just one, but of around half a dozen mathematical texts, including Newton's \textit{Principia}, is a claim hard to accept on the basis of currently available evidence.

Whoever continues to endorse undocumented claims about Tafazzul's translation output should be reminded of what Sir William Jones wrote in a different context \cite[p.~738]{Jones1970Letters}: ``that Moses wrote any of the psalms, may be true; but ought not to be roundly asserted without proof''.

\section*{List of appendices}
\begin{enumerate}[label=\Alph*.]
\item Remarks on transliteration
\item Sh\={u}shtar\={\i}'s account of Tafazzul's life
\item A closer look at  the astronomy in \textbf{\textit{Tuhfa}}
\item Scrutiny of \textbf{\textit{Pamphlet}}
\end{enumerate}
\appendix

\section{Remarks on transliteration}
\label{appendix:translit}

The substitutions for consonants used by F. Steingass in his monumental \textit{Persian-English Dictionary} \cite{Steingass1977Persian} will be adopted here, except when his choices clash with the currently prevalent practice, or when typographical considerations tilt the choice in favour of a different symbol; for the readers' convenience, all departures will be explicitly pointed out. 

Steingass has used five ``compounds, namely \textit{ch} for {\seturdu\novocalize{\<^c>}}, \textit{\underline{gh}} for {\seturdu\novocalize{\<.g>}}, 
\textit{\underline{kh}} for {\seturdu\novocalize{\<_h>}}, \textit{sh} for {\seturdu\novocalize{\<^s>}}, \textit{zh} for {\seturdu\novocalize{\<^z>}}''; these symbols will be replaced here by \textit{\v{c}}, \textit{\v{g}}, \textit{\v{k}}, \textit{\v{s}}, and \textit{\v{z}}, respectively.

Seven other letters carry diacritical points: 
\textit{\'{s}} = {\seturdu\novocalize{\<_t>}},  
\textit{\d{h}} = {\seturdu\novocalize{\<.h>}},
\textit{\'{z}} = {\seturdu\novocalize{\<_d>}}, 
\textit{\d{s}} = {\seturdu\novocalize{\<.s>}}, 
\textit{\d{z}} = {\seturdu\novocalize{\<.d>}},  
\textit{\d{t}} = {\seturdu\novocalize{\<.t>}},  and
\textit{\.{z}} = {\seturdu\novocalize{\<.z>}}. 
 (Steingass: \textit{\td{s}} = 
{\seturdu\novocalize{\<_t>}}, 
\textit{\underline{z}} = {\seturdu\novocalize{\<_d>}}, and 
\textit{\td{t}} = {\seturdu\novocalize{\<.t>}}).

When it occurs at the beginning of a word or syllable, {\seturdu\novocalize{\<n>}} is sounded like \textit{n}; at the end of a word or syllable, if preceded by a long vowel, it has a soft nasal sound like that of \textit{n} in the French word \textit{gar\c{c}on}; when followed by the labials  {\seturdu\novocalize{\<b>}}, {\seturdu\novocalize{\<p>}}, {\seturdu\novocalize{\<f>}}, it assumes the sound of \textit{m}. Steingass's choice for the latter case, namely \textit{\d{m}}, will be respected; for example, {\seturdu\novocalize{\<dunbAlaH>}} (\textit{tail}) will be transliterated as \textit{du\d{m}b\={a}la}, not \textit{dunb\={a}la}. The nasal \textit{n}, which is not treated as a special case by Steingass, will be represented by \~{n}.

In Urdu (but not in Persian), {\seturdu\novocalize{\<h>}} (the \textit{two-eyed h\={e}} has been used, since about 1880, only as the final part of signs showing aspirated sounds (which are not found in Persian). A prime placed next to a consonant will represent aspiration of the corresponding sound; with this convention {\seturdu\novocalize{\<bh>}} $ =$b\textprime{} , {\seturdu\novocalize{\<^ch>}} =	\v{c}\textprime{}, etc. The Urdu letters 
{\seturdu\novocalize{\<,t>}}, {\seturdu\novocalize{\<,r>}} and {\seturdu\novocalize{\<,d>}} will be represented by placing a circle above the corresponding Persian letter; thus {\seturdu\novocalize{\<,t>}} = \r{t}, and so on.

\section{Sh\={u}shtar\={\i}'s account of Tafazzul's life}
\label{appendix:TafBio}

English translations of several passages from \textbf{\textit{Tuhfa}} are sprinkled throughout William Dalrymple's \textit{White Mughals} \cite{Dalrymple2003White}; some authors have used such ready-made excerpts as source material, if not quite as a back-door entry into \textbf{\textit{Tuhfa}}, presumably because they are unable to read Persian. This is a precarious strategy, because Dalrymple himself thanks a Bruce Wannell for ``wonderful translations from the Persian'' (page xxxiv). The translator, though admirably competent, has a habit of hopping over chunks of Sh\={u}shtar\={\i}'s text without any warning, and has good reasons for doing so (see below). \par

In this appendix, I will comment on what Sh\={u}shtar\={\i}, who has been called ``Tafazzul's friend and biographer'' \cite{Schaffer2009Astro}, wrote about Tafazzul's life and works (in the five and a half pages devoted to the topic). Since the assessment presented below is based on an English rendering of a few Persian passages, some preliminary comments on the difficulty of translating a Persian text of that era are necessary. 

First: not only are vowel marks and the \textit{ezafe} (transliterated here as \textit{\'{e}}) omitted, but also the extra slanting stroke that distinguishes \textit{g\={a}f} ({\seturdu\novocalize{\<g>}}\hspace{2pt}) from \textit{k\={a}f} ({\seturdu\novocalize{\<k>}}\hspace{2pt}) is dropped routinely; these parsimonious steps are consistent with the supposition---flattering to some but frustrating to others---that only a highly literate adult will read the text, and such a reader needs no scribal crutches. Second: marks of punctuation were unknown when \textbf{\textit{Tuhfa}} was written. Third: the Persian script has no equivalent of the cases (upper and lower) for each letter. Fourth: no italic or bold fonts, properly so called, were available for adding emphasis in a typical manuscript page. 

No intelligent author writing in Persian could have been unaware of the aforementioned limitations, and various devices were employed to mitigate the difficulty caused by the lack of visible segmentation in the text. The most important among these devices is the symbol {\seturdu\novocalize{\<w>}} when it is used a \textit{word} meaning ``and''. A second device, both a help and a hindrance, is the use of rhyming; helpful because a sentence cannot end before the rhyming pair (or triplet etc.) is completed, but a hindrance for a translator because often the individual words in the pair happen to be synonymous or superfluous. Like most of his contemporaries, Sh\={u}shtar\={\i} was prone to excessive use of rhyming.

The stage has now been set for reading (in translation) Sh\={u}shtar\={\i}'s ``biography'' of Tafazzul, which starts in the middle of p.~442, and ends on p.~447. I will proceed in two steps. An overliteral translation of the relevant text on p.~442 will be displayed first, using a mock Persian style, in that the translated passage will contain no \textit{conventional} punctuation marks, no upper case letters, and---apart from one exception---no italics. Rhyming in the original text will be made visible by transcribing the phrases in italics; the meanings of the phrases will be provided in footnotes immediately after the end of the passage.   In order to enhance the intelligibility of the English version, I have added, within square brackets, the equivalent of what is implicit in the Persian original, and also inserted a small open circle ($\circ$) as an indication for a short or intermediate pause. 

\medskip
\leftskip=20truept
{\small\noindent
[tafa\d{z}\d{z}ul] is among the dignitaries of the capital city lahore \& his auspicious birth took place in that famous city \& afterwards without soliciting the post [he was] appointed as the ambassador representing naww\={a}b \={a}\d{s}af ud-daula ya\d{h}\={a} kh\={a}n$\circ$ sovereign of the entire province of awadh \& lucknow$\circ$ stationed at [the headquarters of] the English government$\circ$ [he was] one of the pre-eminent \textit{fu\d{z}al\={a}y\'{e} n\={a}md\={a}r}$^{1a}$ \& the doyen of  \textit{\d{h}ukam\={a}y\'{e} rozg\={a}r}$^{1b}$ \& was$\circ$ in all branches of knowledge$\circ$ a \textit{f\={a}\d{z}il\'{e} b\={\i}-na\.{z}\={\i}r}$^{2a}$  \& \textit{\textrevglotstop all\={a}ma\'{e} ta\d{h}r\={\i}r}$^{2b}$  especially in \textit{\d{h}ikmiyy\={a}t} \& \textit{il\={a}hiyy\={a}t} (= philosophy \& divinology)$\circ$ plato of the epoch \& aristotle of the age$\circ$ abided awhile in \v{s}\={a}hjah\={a}n-\={a}b\={a}d under the tutelage of contemporary scholars \& in ban\={a}ras$\circ$ from the words of [in the company of] \textit{failsuf\'{e} a\textrevglotstop zam}$^{3a}$ \& \textit{im\={a}m\'{e} akram}$^{3b}$ \textit{\v{s}ai\v{k}\'{e} ajjal} [= \v{s}ei\v{k} the magnificent],  \v{s}ei\v{k} \textrevglotstop al\={\i} \d{h}az\={\i}n$\circ$ acquired knowledge to a high level \&  attained an exalted status$\circ$ his speech-making \& exposition [made him] the envy of the \textit{\^{c}ah-\^{c}aha\'{e} bulbul\'{e} hazar d\={a}st\={a}n dar bah\={a}r\={a}n}$^{4a}$ \& an inimitable model for the \textit{qah-qaha\'{e} kubk\'{e} dar\={\i} dar kohs\={a}r\={a}n}$^{4b}$$\circ$ his open nature was like [p.~442 ends here.] \par

\noindent
$1a$: renowned scholars; $1b$: scholars of the age; $2a$: unrivalled master; $2b$: superb scholar-writer; $3a$: the great philosopher; $3b$: noble leader; $4a,b$: laughter of two song birds, each a familiar motif in Persian poetry.
}

\leftskip=0truept
\medskip

Only Ab\={u} \d{T}\={a}lib \cite{Hoey1885AsafuD} and Sh\={u}shtar\={\i} give Lahore as Tafazzul's place of birth; other authors name Si\={a}lk\={o}\r{t}  ({\seturdu\novocalize{\<siyAlkw,t>}}\hspace{2pt}) instead. Also, Sh\={u}shtar\={\i} is the only one who mentions \v{S}hei\v{k} \textrevglotstop Al\={\i} \d{H}az\={\i}n as one of Tafazzul's teachers. SMA notes both discrepancies; concerning the place of birth, he writes that his elders confirmed that, as stated in \textit{\textrevglotstop Im\={a}d al-Sa\textrevglotstop\={a}dat} \cite{GhulamAli1864Imad}, Tafazzul was born in Si\={a}lk\={o}\r{t}; about Tafazzul receiving some instructions from \textrevglotstop Al\={\i} \d{H}az\={\i}n in Ban\={a}ras, SMA neither confirms nor denies it, saying merely ``It is possible''.

We move now to p.~443. As before, the English version below will not use standard puctuation marks, because the actual text uses none (see Figure~\ref{fig:1}). An additional symbol (\textdoublevertline) will be needed as a substitute for an em dash (---). Since this translation will be compared with another rendering, upper case letters will no longer be shunned, and bold font will be used for some words, so that they may serve as signposts and facilitate the comparison. Speaking of Tafazzul, Sh\={u}shtar\={\i} continues:

\medskip
\leftskip=20truept
{\small\noindent
His open nature was like morning bursting forth$\circ$ \textit{n\={u}r \={a}g\={\i}n} (drenched in light) \& repository of \textit{\textrevglotstop ul\={u}m\'{e} awwal\={\i}n \& \={a}\v{k}ir\={\i}n} [= sciences alpha to omega] \& a most devout and \textbf{pious} Sh\={\i}\textrevglotstop a \& his face illumined by the light of the authority of the holy Im\={a}ms$\circ$ may All\={a}h's blessings be upon them all$\circ$ his well-focussed mind \& its agility to transfer [its contents to others] like a sharp-edged sword \& his munificent attitude$\circ$ manifest \& hidden$\circ$ combined to form an agreeable whole \& in this entire country the light of his beneficence was made patent by the succour he provided to those who had no friends \& to the gatherings of western sages and people with good breeding he was the life and soul [treated] with dignity \& respect \& the truth is that his virtues and excellence have placed him on a high pedestal$\circ$ A whole lifetime and a reed bed [for making pens] is necessary for recording even a soup\c{c}on of his endowments$\circ$ Arabic \& Persian \& English \& R\={u}m\={\i}\textdoublevertline the scholarly language different from the vernacular \& any European who desires to write a book uses this language \& they call it \textbf{Latin}, and the status it enjoys in Europe is similar to that accorded by \textbf{non-Arab} scholars to Arabic\textdoublevertline \& \textbf{Greek} [he] spoke read and wrote well \& for this reason he had \textbf{translated} several scholarly books of Europe into Arabic \& had composed his own books as well \& the most noteworthy among which are a commentary on the Conics of  \textit{\={E}loni\={u}s} ({\seturdu\novocalize\<aylwniyws>})  \& compiled two tracts on \textbf{algebra} one containing algebraic solutions \& one embracing algebraic \& geometric solutions \& [wrote] commentaries on the Conics of \textit{Deopit\={a}l} ({\seturdu\novocalize\<dywpitAl>}\hspace{2pt}) and the Conics of Simson \& through debating \& book-reading wrote so many marginalia and glosses on books of \d{h}ad\={\i}th \& \textbf{jurisprudence} of the two sects \& Islamic philosophy \& other [?] \textrevglotstop ul\={u}m [sciences, or branches of knowledge] that even a fraction [end of p.~443]. \par
}

\leftskip=0truept
\medskip

As it happens, Dalrymple has also quoted, in a long footnote on p.~271 of \textit{White Mughals} \cite{Dalrymple2003White}, an excerpt from the pages of \textbf{\textit{Tuhfa}}. Tafazzul is described as a 

\medskip
\leftskip=20truept
{\small\noindent
\textbf{pious} Shiite, who also knew, apart from Persian and Arabic, English and the Roman tongue which they call Latin, which is the learned tongue of the Europeans in which they write their scholarly books, and which has the same position among them as Arabic among \textbf{non-Arab} Muslims. Tafazul even knew \textbf{Greek} and had \textbf{translated several} books by European scholars into Arabic, apart from his own writings on \textbf{algebra} and \textbf{jurisprudence}. [next page begins] India should be proud to have brought forth such a scholar \ldots\ however much his position gave him the attributes of wealth and status, he never changed his courteous and egalitarian behaviour towards the poor and the weak’; Seyyed Abd al-Latif Shushtari, \textit{Kitab Tuhfat al-\bsq Alam} (written Hyderabad, 1802; lithographed Bombay, 1847), p.450 [sic].\par
}

\leftskip=0truept
\medskip

The reader will be able to see what (and how much) is missing from the \textit{White Mughals} translation, and perhaps will also be able to understand why the translator decided to lighten his (own and the reader's) burden. 

What Sh\={u}shtar\={\i} wrote about the status of Latin among European scholars was true once, but not at the beginning of the nineteenth century, when \textbf{\textit{Tuhfa}} was written. Galileo had already perceived, when he published his famous \textit{Dialogo} in 1632, the advantages of writing in vernacular Italian rather than scholarly Latin. \textit{Optics}, Newton's second book, was written in pellucid English. Marquis de l'Hospital's text on differentials was, like his text on conics, in French. For better or for worse, Latin started losing ground in the early decades of the eighteenth century, so much so that d'Alembert was alarmed \cite{d'Alembert1995Preliminary}:

\medskip
\leftskip=20truept
{\small\noindent
The practice today of writing everything in the vulgar tongue has doubtless contributed to strengthening this prejudice [against the ancients], and perhaps is more pernicious than the prejudice itself. Since our language was spread throughout all Europe, we decided that it was time to substitute it for Latin, which had been the language of our scholars since the renaissance of letters. I acknowledge that it is much more excusable for a philosopher to write in French than for a Frenchman to make Latin verses. I would be willing even to agree that this practice has contributed to making enlightenment more general, if indeed broadening the outer surface also broadens the mind within. However, an inconvenience that we certainly ought to have foreseen results from it. The scholars of other nations for whom we have set the example have rightly thought that they would write still better in their own language than in ours. Thus England has imitated us. Latin, which seemed to have taken refuge in Germany, is gradually losing ground there. I have no doubt that Germany will soon be followed by the Swedes, the Danes, and the Russians. Thus, before the end of the eighteenth century, a philosopher who would like to educate himself thoroughly concerning the discoveries of his predecessors will be required to burden his memory with seven or eight different languages. \par
}

\leftskip=0truept
\medskip

\section{ A closer look at the astronomy in \textbf{\textit{Tuhfa}}}
\label{appendix:CloseTuhfa}

The word {\seturdu\novocalize{\<.hakIm>}} (\textit{\d{hak\={\i}m}}) and its pulural {\seturdu\novocalize{\<.hukamA'>}}  (\textit{\d{hukam\={a}\textglotstop}}) occur rather frequently in \textbf{\textit{Tuhfa}}. The singular form may be translated roughly as \textit{sage}, \textit{learned}, \textit{wise}, \textit{doctor}, \textit{physician}, \textit{scientist}, etc. On p.~352  a new section begins, whose heading (given in the margin) may be rendered as follows: \textit{opinions of European sages about the positions and movements of fixed stars and planets}. A translation of the first six lines is given below:

\medskip
\leftskip=20truept
{\small
The luminous Sun provides light to the fixed stars and the planets, benefitting thereby \textit{all the worlds} ({\seturdu\novocalize{\<bijamI` `awAlim>}}) and providing sustenance to all. It remains stationary in the centre of the planetary orbits, while all the other bodies describe orbits around it, and acquire light from it. It does not move [from its position] but spins around its axis from the west to east. The (globe of the) Earth is counted as one of the planets and it goes around the sun, at a distance of forty five crores and fifteen lac miles (451.5 million miles) from the sun, and the sun is ten crores and two lac (100.2 million) times bigger than the Earth. What appears to be the rising of the Sun from the east is an error of perception \ldots\par
}
\leftskip=0truept
\medskip
Sh\={u}shtar\={\i}'s ideas about the fixed stars are as baffling as his cavalier attitude towards numbers (with four significant figures), and his omission of the source of the data mentioned by him. Numbers, when handled by someone who is not numerate, signify nothing. To take account of the uncertainties of observations, and to eliminate the need of a calculator in the discussion, let us round off the Sun-to-Earth distance to 450 million miles, and the relative size of the Sun to 100 million. Our first task is to find data that would have been considered reliable in his time.

The required information was available in \textit{A New Theory of Earth}, a book written by William Whiston (1667-1752)---student and friend of Newton, and his successor to the Lucasian chair of mathematics in Cambridge, but a marginalized figure during the last two decades of his life. First published in 1696, his book went through six editions, the last in 1755. Whiston gives the mean distance of the Earth  from the Sun as 81 million miles; the diameters of the Sun and the Earth, as 763000  and 7970 miles respectively. This means that the diameter of the Sun (relative to that of the Earth) is about 100 (not 100 million). To give Sh\={u}shtar\={\i} the benefit of the doubt, let us define the relative size of the Sun to be the ratio of the \textit{volumes} of the two bodies; the relative size of the Sun now comes out to be $100^3$, or 1 million, still very much smaller than the number quoted by Sh\={u}shtar\={\i}. 

\end{multicols}

\noindent\rule{\textwidth}{0.5pt}
\vspace{-12pt}
\begin{figure}[ht]
\begin{minipage}[b]{0.45\linewidth}
\centering
\includegraphics[width=\textwidth]{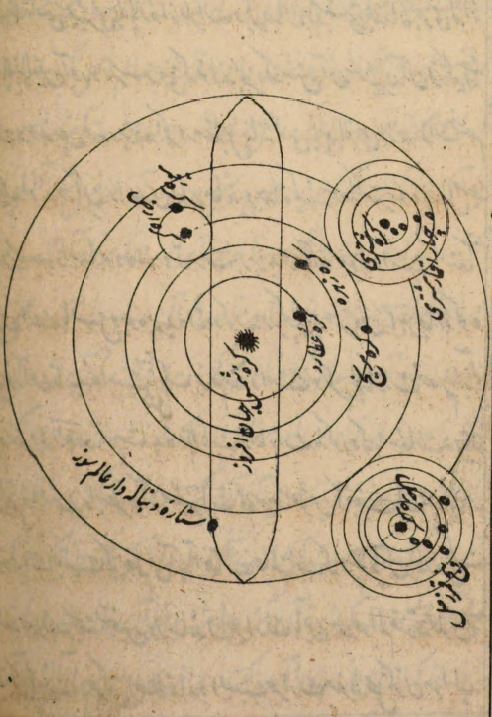}
\caption{The solar system as depicted in \textbf{\textit{Tuhfa}} \cite{Shushtari1847Tuhfa}.}
\label{fig:4}
\end{minipage}
\hspace{10mm}
\begin{minipage}[b]{0.45\linewidth}
\centering
\includegraphics[width=0.92\textwidth]{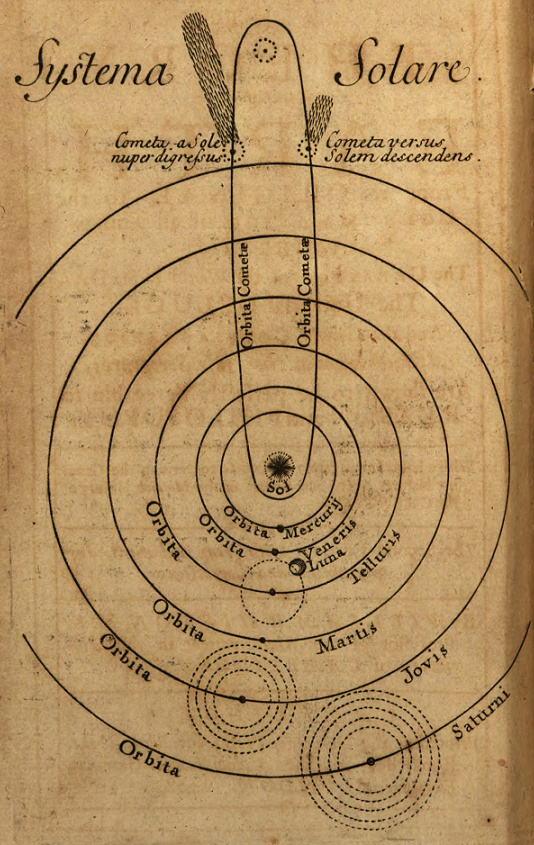}
\caption{The solar system as sketched by Whiston in \textit{A New Theory of the Earth} \cite{Whiston1708New}.}
\label{fig:5}
\end{minipage}
\vspace{-5mm}
\end{figure}

\noindent\rule{\textwidth}{0.5pt}

\begin{multicols}{2}

Sh\={u}shtar\={\i} was fascinated by the fact that Newton's law of gravitational attraction was able to account not only for the motion of the planets but also for the behaviour of the comets, which had previously been regarded as vagrant bodies coming in and out of our view at random intervals. He showed his enthusiasm by inserting a figure---the only illustration in a book of some 600 pages---that portrays the Newtonian conception of the solar system. It will be helpful to read, before examining this figure, a modern translation of how Newton described the arrangement of celestial bodies in our solar system \cite[p.~586]{Newton1999Principia}: 

\medskip
\leftskip=20truept
{\small
The six primary planets revolve about the sun in circles concentric with the sun, with the same direction of motion, and very nearly in the same plane. Ten moons revolve
about the earth, Jupiter, and Saturn in concentric circles, with the same direction of motion, very nearly in the planes of the orbits of the planets. And all these regular
motions do not have their origin in mechanical causes, since comets go freely in very eccentric orbits and into all parts of the heavens. And with this kind of motion the
comets pass very swiftly and very easily through the orbits of the planets; and in their aphelia, where they are slower and spend a longer time, they are at the greatest possible distance from one another, so as to attract one another as little as possible.\par}

\leftskip=0truept
\medskip

Figure~\ref{fig:4} reproduces the model of the solar system presented in \textbf{\textit{Tuhfa}} \cite[p.~360]{Shushtari1847Tuhfa}. One finds five sun-centred circles in the figure, though one would expect to see six, one for each planet. That Mars and the Earth have been placed in the same orbit is clearly an oversight (not necessarily on the author's part), but the discrepancy is unlikely to mislead a reader who has learnt some astronomy. That the closed, non-circulartrajectory, supposed to show the path of a comet, is also centred at the sun betrays Sh\={u}shtar\={\i}'s utter misapprehension of cometary orbits.  Figure~\ref{fig:5} is taken from \textit{A New Theory of Earth} \cite{Whiston1708New}, where it appears as the frontispiece.

Apart from the comet, the names of all other bodies shown in Figure~\ref{fig:4} begin with the word {\seturdu\novocalize{\<kuraH>}}, which may be translated in the present context as \textit{spherical body}; since this word serves no useful purpose, it will be dropped without further comments.  The Sun is called  {\seturdu\novocalize{\<^sams jahAn afrwz>}}, meaning \textit{Sun, the world-illuminator}.  The Persian word for a comet is \textit{sit\={a}ra\'{e} du\d{m}b\={a}la-d\={a}r} (meaning \textit{star with a tail}), but  the more elaborate label \textit{sit\={a}ra\'{e} du\d{m}b\={a}la-d\={a}r\'{e} \textrevglotstop\={a}lam s\={u}z} (\textit{world-scorching tailed star}) is used in the figure. If we allow for the author's fondness for verbosity, the epithet applied to the Sun no explanation, but the relevance of the descriptive tag attached to the comet might not be equally obvious to the average reader.
 
The heart of Whiston's new theory was the role played by comets. Being an ardent admirer of Newton, Whiston made free use of the scriptures and of Newton's concept of gravitational attraction and his theories of comets to concoct a unified theory of the creation---as well as the destruction---of the Earth. The scope of the book can be more easily grasped if we look at the complete title: \textit{A New Theory of the Earth, from Its Original, to the Consummation of All Things: Wherein the Creation of the World in Six Days, the Universal Deluge, and the General Conflagration, as Laid Down in the Holy Scriptures, are Shewn to be Perfectly Agreeable to Reason and Philosophy}. For a while the book became a best seller and was highly praised by many, including John Locke \cite{KRN2015CSCB2I,Strauss2016Whiston}.  This theory gets a warm reception in Tuhfa, and Sh\={u}shtar\={\i}. 

Transcribing \textit{Millennium} as {\seturdu\novocalize{\<milywn>}}  (\textit{mily\={o}n}), Sh\={u}shtar\={\i} writes (p.~354): ``a comet, because of its close approach to the sun during its travel, becomes extremely hot, and if [in that state] it collides with a planet, especially the Earth, the planet is scorched; this is how \textit{qay\={a}mat}, which is called \textit{mily\={o}n}, is to be interpreted; the whole world is incinerated, and not a single living being (man or beast) is left, nor any trace of minerals or vegetation; this view is at variance with that held by the ancient sages, who did not believe in  \textit{qay\={a}mat}.'' Sh\={u}shtar\={\i} does not mention Whiston, but it is hard to think of any other author who spoke so much, and with such passion, about comets and the Millennium \cite{KRN2015CSCB2I}. Whiston's theory did not exert any lasting influence in Britain, and he had become, already in his lifetime, the butt of many a satirist. One of Sh\={u}shtar\={\i}'s Company friends must have introduced him to Whiston's theory, and might even have shown him Figure~\ref{fig:5} in a copy of \textit{A New Theory of the Earth}. This conjecture would also explain the great resemblance between Figures~\ref{fig:4} and \ref{fig:5}, and might also be one of the many details whose loss is lamented by Dalrymple \cite[p.~376]{Dalrymple2003White}: ``In a similar manner, although the exact details are now sadly lost, James and Abdul Lateef Shushtari seem to have been spending their nights on the
Residency roof, busy comparing notes to see how Indian, Islamic and European astronomical systems could be reconciled, and what each could learn from the other.''

Whiston died long before the announcement, in 1781, of a new planet, that came to be called Uranus, but \textbf{\textit{Tuhfa}} was written more than twenty years after the discovery. One might expect that an author like Sh\={u}shtar\={\i}, who wanted to introduce the new astronomy to his readers, would not omit the most recently discovered planet, but the news had either not reached him or he did not consider it important enough to deserve a mention. 

Readers who have followed thus far do not need more examples to form their estimates of the scientific part of Sh\={u}shtar\={\i}'s book. For my part, I believe that, since dilettantism is not an asset for someone who wants to popularize science, Sh\={u}shtar\={\i} should have set himself the humbler goal of acquiring a sound knowledge of astronomy and mathematics before filling the pages of \textbf{\textit{Tuhfa}} with information he was unable to acquire or digest first-hand.  

\end{multicols}

\noindent\rule{1.03\textwidth}{0.5pt}
\vspace{-12pt}
\begin{figure}[ht]
\begin{minipage}[b]{0.33\linewidth}
\centering
\includegraphics[width=\textwidth]{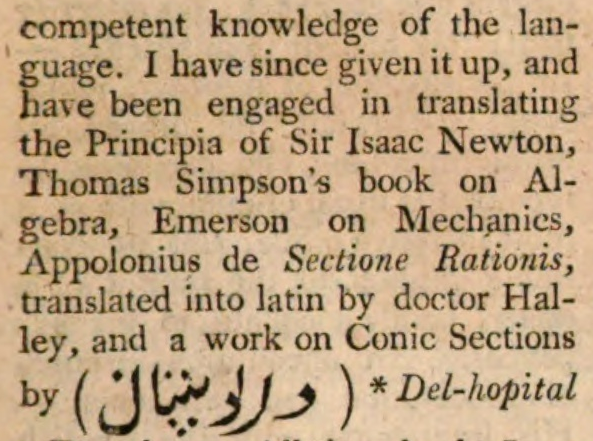}
\caption{An excerpt from p.~7 of \textbf{\textit{Obituary}} showing the segment where the name \textit{de l'Hospital} (misspelt) is written in Persian; the rhombus ($\smallblackdiamond$, a diacritical point) at the extreme left is an obvious printing error.}
\label{fig:6}
\end{minipage}
\hspace{3mm}
\begin{minipage}[b]{0.67\linewidth}
\centering
\includegraphics[width=0.92\textwidth]{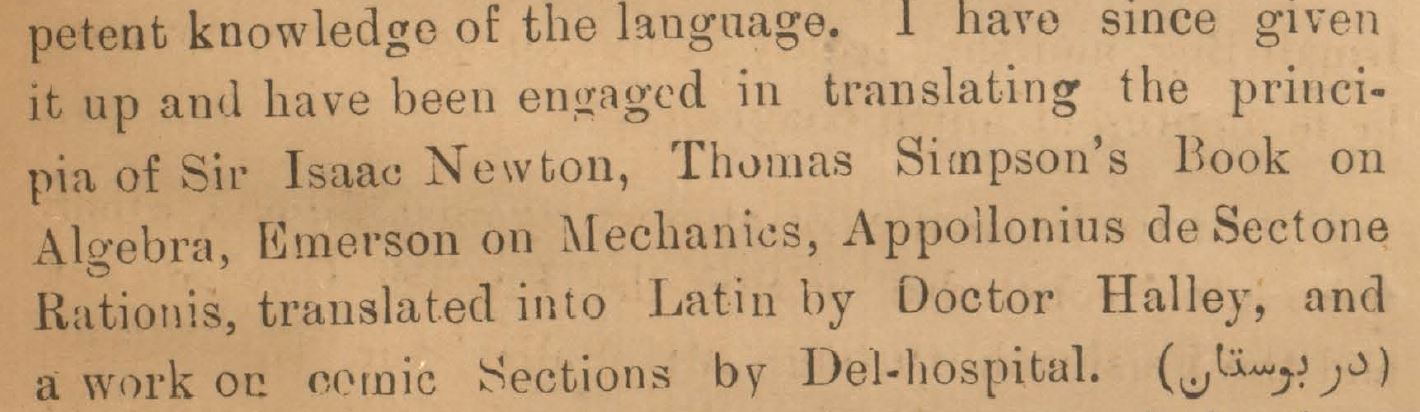}
\caption{This excerpt from p.~6 of \textbf{\textit{Leaflet}} (Part II) should be compared with the text in Figure~\ref{fig:6} to appreciate the nature of corruptions which accompanied the typesetting of \textbf{\textit{Leaflet}}. To give a few examples: ``principia'' for ``Principia'', ``Sectone'' for ``\textit{Sectione}'', ``comic Sections'' for ``Conic Sections'', and ``Del-hospital'' for ``\textit{Del-hopital}''. The Persian text ({\seturdu\novocalize\<darbwstAn>}) now bears no relation to the French name, because it stands for a supposedly corrected version of the Persian text appearing in \textbf{\textit{Obituary}} (see  Fig.~\ref{fig:6}).}
\label{fig:7}
\end{minipage}
\vspace{-10mm}
\end{figure}

\noindent\rule{1.03\textwidth}{0.5pt}

\begin{multicols}{2}

\section{Scrutiny of  \textbf{\textit{Pamphlet}}}
\label{appendix:ScrutPamp}

It has been stated earlier (\S~\ref{sec:sources}) that the relevant parts (namely I and II) of \textbf{\textit{Leaflet}} are meant to be mere copies (but in a different format) of documents that had been published earlier, and that the text in each part is replete with typesetting errors. The purpose of this appendix is to substantiate this remark by showing snippets from  \textbf{\textit{Obituary}} and Part II of \textbf{\textit{Leaflet}}. Figures~\ref{fig:6} and \ref{fig:7} compare short segments from the two sources. 

It is worth recalling that Anderson had sent to Campbell a translation of Tafazzul's Persian letter. The names (in Latin characters) of British authors and the titles of their works escaped mutilation, but Figure~\ref{fig:6} shows that Persian script was used, at least in \textbf{\textit{Obituary}}, for writing \textit{de l'Hospital}, the name (misspelt as \textit{Del-Hospital}) of the celebrated mathematician who wrote the first textbook on the calculus of differentials; those who are able to read Persian will see that an unwarranted diacritical mark (a rhombus, usually called a ``dot'') has appeared on the last letter. Whoever typeset (or edited) the text for \textbf{\textit{Leaflet}} failed to realise that the dot was spurious and took the text to be an intelligible Persian word (instead of a foreign name), that the desideratum was a Persian transcription of \textit{Del-Hospital}, and ``corrected'' the Persian text to {\seturdu\novocalize\<darbwstAn>} (see Figure~\ref{fig:7}), which in intelligible, but a red herring (or a complete mess).

\end{multicols}
\end{document}